\documentclass[11pt]{article}
\usepackage{jcapmod}

\usepackage{booktabs}
\usepackage[english]{babel}
\usepackage{amsmath,amssymb,amsbsy,amstext, amsthm, simplewick}
\usepackage{hyperref}
\usepackage{graphicx}
\usepackage{amsfonts}
\usepackage{amssymb}
\usepackage[small]{caption}
\usepackage{upgreek}
\usepackage{wrapfig}
 \usepackage{exscale,relsize}

\newcommand{\otoprule}{\midrule[\heavyrulewidth]}
\usepackage{colortbl}
\definecolor{lightgreen}{cmyk}{0.2, 0, 0.2, 0.2}
\definecolor{lightgray}{cmyk}{0.1,0.2,0,0.1}
\definecolor{lightgray2}{cmyk}{0.1,0.1,0,0.1}

\makeatletter
\newlength{\apb@width}
\newcommand{\autoparbox}[2][c]{\settowidth{\apb@width}{#2}\parbox[#1]{\apb@width}{#2}}
\newcommand{\includegraphicsbox}[2][]{\autoparbox{\includegraphics[#1]{#2}}}
\makeatother

\numberwithin{equation}{section}

\def\beq{\begin{equation}}
\def\eeq{\end{equation}}

\def\bea{\begin{eqnarray}}
\def\eea{\end{eqnarray}}

\def\d{{\rm d}}

\def\dalpha{{\dot \alpha}}
\def\Df{{\bar D}}

\def\Db{\bar{{\cal D}}}

\def\dalpha{\dot{\alpha}}

\def\beq{\begin{equation}}
\def\eeq{\end{equation}}
\def\bea{\begin{eqnarray}}
\def\eea{\end{eqnarray}}

\def\Mp{M_{\rm pl}}
\def\d{{\rm d}}

\def\G{\Phi}
\def\Gd{\Phi^{\dagger}}

\def\Phid{{\Phi^{\dagger}}}
\def\phid{{\bar \phi}}

\DeclareRobustCommand{\SkipTocEntry}[4]{}

\begin{document}

\begin{titlepage}

\setcounter{page}{1} \baselineskip=15.5pt \thispagestyle{empty}

\bigskip\

\vspace{2cm}
\begin{center}
{\fontsize{20}{32}\selectfont  \bf Signatures of Supersymmetry\\ \vskip 10pt from the Early Universe}
\end{center}

\vspace{0.5cm}
\begin{center}
{\fontsize{14}{30}\selectfont   Daniel Baumann$^{\clubsuit,\heartsuit}$ and Daniel Green$^\clubsuit$}
\end{center}


\begin{center}
\vskip 8pt
\textsl{$^\clubsuit$ School of Natural Sciences,
 Institute for Advanced Study,
Princeton, NJ 08540, USA}

\vskip 7pt
\textsl{$^\heartsuit$ Centre for Theoretical Cosmology, Cambridge University, Cambridge, CB3 0WA, UK}

\end{center}

\vspace{1.2cm}
\hrule \vspace{0.3cm}
{ \noindent \textbf{Abstract} \\[0.2cm]
\noindent 
Supersymmetry plays a fundamental role in the radiative stability of many inflationary models.
Spontaneous breaking of the symmetry inevitably leads to fields with masses of order the Hubble scale during inflation.
When these fields couple to the inflaton they produce a unique signature in the squeezed limit of the three-point function of primordial curvature perturbations.
In this paper, we make this connection between naturalness, supersymmetry, Hubble-mass degrees of freedom and the squeezed limit precise.
To study the physics in a model-insensitive way, we develop a supersymmetric effective theory of inflation.
We use the effective theory to classify all possible interactions between the inflaton and the additional fields, and determine which ones naturally allow large non-Gaussianities when protected by supersymmetry.
Finally, we discuss the tantalizing prospect of using cosmological observations as a probe of supersymmetry.}
 \vspace{0.3cm}
 \hrule

\vspace{0.6cm}
\end{titlepage}

\tableofcontents

\newpage
\section{Introduction}

Supersymmetry (SUSY)~\cite{SUSY} is a powerful way to protect scalar fields from quantum corrections.
If the symmetry is unbroken, it controls dangerous radiative effects by enforcing exact cancellations between boson and fermion loops. Even if the supersymmetry is broken at low energies---as it has to be if it is to describe the real world---the appearance of supersymmetry at high energies can still help to regulate loop effects. 
In inflation~\cite{inflation}, SUSY is often required to achieve technical naturalness \cite{'tHooft:1979bh} of the quantum-corrected inflaton action~\cite{ArkaniHamed:2003mz, Lyth:1998xn}.
Hence, even if SUSY is not discovered at the weak scale, naturalness motivates SUSY in inflation.
Can we use cosmological experiments to probe supersymmetry even if it is out of reach of particle colliders?
What are generic signatures of supersymmetry during inflation?
In this paper we will address these questions.

\vskip 6pt
An inevitable consequence of SUSY during inflation is the presence of additional scalar fields, with masses of order the Hubble scale $H$.
The appearance of the Hubble scale as a preferred mass scale is robust and simply related to fact that the size of spontaneous SUSY breaking during inflation is determined by the vacuum energy, $\langle {\cal F}_X \rangle^2 = 3\Mp^2 H^2$. 
Gravity mediates this breaking to the scalar sector, leading to the relation
\beq
m \sim \frac{\langle {\cal F}_X \rangle}{\Mp} \sim H\ .
\eeq
This effect is familiar from the supergravity (SUGRA) eta problem~\cite{Copeland:1994vg}, where the induced mass for the inflaton field, $m_\phi^2 \sim H^2$, threatens to end inflation prematurely.
Inflation will only last for at least 60 $e$-folds if the inflaton mass is tuned to be smaller than $H$, or if a smaller mass is protected by an additional global symmetry.
This makes it clear that additional fields without protective global symmetries will inherit Hubble scale masses from SUSY breaking.
It is less clear that these fields can have observational consequences.  

\vskip 6pt
We are looking for an observable that is sensitive to the number of light degrees of freedom during inflation, and that, ideally, would probe the masses of the additional fields.
A promising candidate is the {\it squeezed limit} of the three-point function of primordial curvature perturbations~$\zeta$\,:
\beq
 \includegraphicsbox[scale=.85]{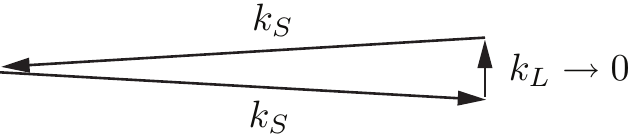} \nonumber
\eeq
This describes the correlation between a long-wavelength fluctuation $\zeta_L = \zeta(k_L)$ and two short-wavelength fluctuations $\zeta_S =\zeta(k_S)$, i.e.~$\zeta_L $ modifies the short-scale power spectrum $\langle \zeta_S^2 \rangle$. 
However, in single-field inflation, constant $\zeta_L$ is a pure gauge mode and hence isn't locally observable. The long-wavelength mode therefore has no effect on the short-wavelength power in the zero momentum limit $k_L \to 0$. Any physical effects are proportional to $\big\{\dot \zeta_L, \nabla^2 \zeta_L \big\} \propto k^2_L \zeta_L$ and hence suppressed by $(k_L/k_S)^2$~\cite{Maldacena:2002vr,Creminelli:2004yq,Creminelli:2011rh}.
On the other hand, if the fluctuations we observe were produced by a second light field, the superhorizon fluctuations of the second field are locally observable at the time when they
are converted to curvature fluctuations. Large non-Gaussianity is then possible in the squeezed limit.
These facts are summarized by the momentum dependence of the three-point function in the squeezed limit 
\beq
\lim_{k_L \to 0} \langle \zeta_{{\bf k}_L}  \zeta_{{\bf k}_{S_1}}  \zeta_{{\bf k}_{S_2}} \rangle \propto \frac{1}{k_L^\alpha}\ .
\eeq
Here, single-field inflation corresponds to $\alpha = 1$ \cite{Creminelli:2011rh}, while
multiple light fields with $m \ll H$ can lead to non-Gaussianity with $\alpha = 3$ \cite{Lyth:2001nq}.\footnote{Here, and in the rest of the paper, we are assuming perfectly scale-invariant fluctuations, $k^3 \langle \zeta_k^2 \rangle \propto k^0$. It is straightforward to restore (percent-level) deviations from scale-invariance, $k^3 \langle \zeta_k^2 \rangle \propto k^{n_s-1}$, in all of our results.} Until recently, no example was known with $\alpha$ different from $1$ or $3$.
We see that measurements of the squeezed limit have the potential to determine whether one or more fundamental fields were relevant for producing the primordial fluctuations.

\vskip 6pt
How does the presence of Hubble-mass degrees of freedom during inflation affect the squeezed limit? 
For the massive `isocurvaton' fields $\sigma$ to have an impact on observables, their fluctuations have to be converted into curvature fluctuations $\zeta$---either during inflation or afterwards.
Recently, Chen and Wang~\cite{Chen:2009zp}  observed that direct couplings between the inflaton $\phi$ and a massive isocurvaton~$\sigma$ lead to a squeezed limit with the following momentum dependence: 
\beq
\alpha = \frac{3}{2} + \sqrt{\frac{9}{4} - \frac{m^2_\sigma}{H^2}}\ . \label{equ:QSFIalpha}
\eeq
These models of {\it quasi-single-field inflation} (QSFI) therefore allow for a scaling in the squeezed limit that is intermediate\footnote{Such intermediate scalings may also be engineered in some curvaton scenarios~\cite{Byrnes:2009pe, Shandera:2010ei}, as well as in theories with non-Bunch-Davies initial states~\cite{Chialva:2011hc}.} between the scaling of single-field and multi-field inflation: $\frac{3}{2} < \alpha <  3$.
The appearance of the mass $m_\sigma$ in eq.~(\ref{equ:QSFIalpha}) has an intuitive explanation: massive fields decay when their wavelengths becomes larger than the Hubble radius. At the time when the short modes cross the horizon, the amplitude of the long mode is therefore suppressed relative to the amplitude when it crossed the horizon. This affects the momentum dependence of the squeezed limit by an amount that depends on $m_\sigma$ (see \S\ref{sec:QSFIreview} for a derivation).

\vskip 6pt
The squeezed limit is both theoretical clean and observationally relevant.
Future cosmic microwave background (CMB) data \cite{Planck} will span a large range of scales and therefore provide access to the squeezed limit of the primordial perturbations.
These measurements will be complemented by large-scale structure (LSS) observations~\cite{SDSS}. 
In particular, in recent years, {\it scale-dependent halo bias}\footnote{Galaxies reside in dark matter halos. Halos are biased tracers of the underlying dark matter density field: $\delta_{\rm h} = b \delta_{\rm m}$. The halo bias $b(k) = \langle \delta_{\rm m}\delta_{\rm h}\rangle/\langle \delta_{\rm m}\delta_{\rm m}\rangle$ develops a characteristic scale-dependence in the presence of non-Gaussianity with a non-trivial squeezed limit of the primordial three-point function. We can hope to detect this effect by measuring the scale-dependence of the galaxy two-point correlation function on large scales.}~\cite{Dalal:2007cu, Matarrese:2008nc} has emerged as a sensitive probe of primordial non-Gaussianity~\cite{Slosar:2008hx, Xia:2011hj}. In this case, the signal is dominated by the squeezed limit of the three-point function~\cite{Schmidt:2010gw}: 
\beq
\Delta b(k) \propto \frac{f_{\rm NL}}{k^{\alpha -1}}\ .
\eeq
The combination of CMB and LSS data therefore suggests the exciting possibility of using observations to probe Hubble-mass degrees of freedom during inflation. As we explained, these fields have their natural home in supersymmetry. In this paper, we will make this relation concrete. Our results will solidify the connection between naturalness, supersymmetry and the squeezed limit of non-Gaussianities. To use cosmological observations to find evidence for a new spacetime symmetry of Nature is, of course, a very tantalizing prospect.

\vskip 6pt
We have two primary goals in this work:
\begin{enumerate}
\item[1)] We wish to develop a broadly applicable framework for incorporating supersymmetry in effective theories of inflation. The starting point for this investigation will be the effective theory of inflationary fluctuations of Cheung et al.~\cite{Cheung}---a theory of Goldstone bosons of spontaneously broken time translations. We will show how the Goldstone fields are embedded in a supersymmetric theory (see also \cite{Senatore:2010wk, Justin}). Along the way we will have to understand how to couple higher-derivative theories to supergravity~\cite{Paper1}, how to cancel tadpoles, and how to characterize the couplings of matter fields to auxiliary supergravity fields. The end product will be the\ {\it supersymmetric effective theory of inflation}. This is an effective theory of the inflationary {\it fluctuations} and therefore complementary to the large body of work on SUSY theories for inflationary {\it backgrounds} (see e.g.~\cite{Lyth:1998xn, Yamaguchi:2011kg}).

\item[2)] We then use this theory for a model-independent description of the interactions between the inflaton and additional Hubble-mass fields. The supersymmetric framework allows us to analyze naturalness in a UV-insensitive way. One of our main goals will be to classify all technically natural SUSY implementations of quasi-single-field inflation.
\end{enumerate}

\vskip 10pt
The outline of the paper is as follows: In Section~\ref{sec:EFT}, we construct a supersymmetric generalization of the effective theory of inflation of Cheung et al.~\cite{Cheung}.
We will illustrate the formalism with two important examples: in Section~\ref{sec:cs}, we derive supersymmetric models of inflation with small sound speed, while in Section~\ref{sec:QSFI}, we present  the first explicit models of QSFI in SUSY.
We classify all possible variations of QSFI and determine which scenarios can lead to naturally large non-Gaussianity when protected by SUSY.
We discuss the observational signatures of these theories.
Sections~\ref{sec:cs} and \ref{sec:QSFI} are self-contained and can therefore be read separately and in any order.
We conclude with Section~\ref{sec:conclusions}.

Two appendices supplement the computations of the main text: In Appendix~\ref{sec:SUGRA}, we show how to couple higher-derivative theories consistently to supergravity. This is a largely non-technical summary of the results of a companion paper~\cite{Paper1}. We use results from this appendix in Section~\ref{sec:cs}.
In Appendix~\ref{sec:QSFI2}, we develop simple estimation techniques to derive the amplitude and the squeezed limit of the bispectrum for models of quasi-single-field inflation. These results are required to reproduce the results of Section~\ref{sec:QSFI}.

\newpage
\section{The Supersymmetric Effective Theory of Inflation}
\label{sec:EFT}

\subsection{Adiabatic Fluctuations as Goldstone Bosons}
\label{sec:adi}

We begin with a quick review of the effective theory of inflation of Cheung et al.~\cite{Cheung} (see also \cite{Creminelli:2006xe}). Readers familiar with this work may skip directly to the next subsection where we generalize the theory to include supersymmetry.

\vskip 6pt
The first step in constructing effective field theories (EFT) is identifying the relevant degrees of freedom for the measurements of interest. Here, we are interested in an EFT that describes CMB observables.
We assume that the origin of the CMB temperatures fluctuations can be traced back to quantum fluctuations during inflation that froze and became classical when their frequencies $\omega$ matched the Hubble expansion rate $H$. The goal is therefore an EFT for the inflationary fluctuations valid at energies $\omega \sim H$.

\vskip 6pt
  \begin{wrapfigure}{r}{.4\textwidth}
  \vspace{.0cm}
\begin{center}
\includegraphics[width=0.35\textwidth]{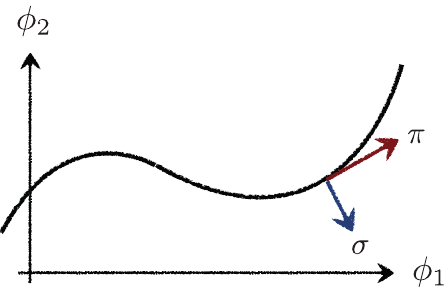}
\end{center}
\vspace{0.5cm}
\end{wrapfigure} 
To construct this EFT, we start from the crucial insight that
time-dependent FRW backgrounds, such as the inflationary quasi-de Sitter spacetimes, spontaneously break time-translation invariance. The relevant degree of freedom of the EFT is then the Goldstone boson associated with the symmetry breaking.  
We introduce the Goldstone mode as a spacetime-dependent transformation along the direction of the broken symmetry, i.e.~we perform a spacetime-dependent time shift
\beq
\varphi \equiv t + \pi(x) \ ,
\eeq
 where $t\to t+ \xi_0(x)$ and $\pi \to \pi - \xi_0(x)$. 
 We note that the field $\pi$ parameterizes adiabatic perturbations, i.e.~perturbations corresponding to a {\it common, local shift in time} for a set of matter fields $\phi_a$, $a=1, \cdots, N$,
 \beq
\delta \phi_a(x) \equiv \phi_a(t+\pi(x)) - \bar \phi_a(t)\ .
\eeq
Moreover, the field $\pi$ is related to the comoving curvature perturbation $\zeta$ by a gauge transformation,
\beq
 \zeta = - H \pi\ .
 \eeq
The low-energy expansion of the field $\varphi(x) = t+\pi(x)$ is the {\it effective theory of inflation}~\cite{Cheung}
\beq
{\cal L}_{\rm eff} = f(\varphi, (\partial_\mu \varphi)^2, \Box \varphi, \cdots)\ .
\eeq
\vskip 4pt
{\it Slow-roll inflation.} \hskip 4pt
At lowest order in derivatives, we have a (time-dependent) potential and a kinetic term for $\varphi$,
\beq
\label{equ:lowest}
{\cal L}_{\rm eff} = - \Lambda(\varphi)  + c(\varphi) (\partial_\mu \varphi)^2 +\cdots
\eeq
To cancel tadpoles (i.e.~terms linear in the fluctuations $\pi$), the coefficients in eq.~(\ref{equ:lowest}) are fixed in terms of the Hubble parameter:
$\Lambda(\varphi) \equiv  \Mp^2(3H^2(\varphi) + \dot H)$ and $c(\varphi) \equiv \Mp^2 \dot H$. This ensures that we are expanding around the correct inflationary background $H(t)$.
Eq.~(\ref{equ:lowest}) is, of course, nothing but slow-roll inflation in disguise
\begin{align}
\label{equ:Lsr}
{\cal L}_{\rm s.r.} = \Mp^2 \dot H (\partial_\mu \varphi)^2  - \Mp^2(3H^2(\varphi) + \dot H) \quad \leftarrow \quad - \tfrac{1}{2}(\partial_\mu \phi)^2 - V(\phi)\ . 
\end{align}
A drastic simplification occurs in the effective action when we take the so-called {\it decoupling limit}.
In this limit we ignore the mixing between the matter fluctuations $\pi$ and the metric perturbations $\delta g^{\mu \nu}$---i.e.~we evaluate the Goldstone action in the homogeneous background spacetime: $g^{\mu \nu} \to \bar g^{\mu \nu}$. 
This approximation only leads to slow-roll suppressed errors in the correlation functions evaluated at horizon-crossing $\omega \sim H$.
Since the effective theory of inflation is most powerful when it describes large non-Gaussianities arising from matter self-interactions, we are justified to work in the decoupling limit---e.g.
\beq
\label{equ:op}
(\partial_\mu \varphi)^2 \equiv g^{\mu \nu} \partial_\mu \varphi \partial_\nu \varphi \ \to \ - 1 - 2 \dot \pi + (\partial_\mu \pi)^2 \ .
\eeq
The action (\ref{equ:Lsr}) then becomes
\beq
{\cal L}_{\rm s.r.} = \Mp^2 \dot H (\partial_\mu \pi)^2\ .
\eeq
We find that $\pi$ is massless and purely Gaussian for slow-roll inflation in the decoupling limit.

\vskip 4pt
{\it Theories with small sound speed.} \hskip 4pt
To go beyond the simple free-field theory, we add higher-derivative terms.
For single derivatives acting on $\varphi = t +\pi$, the leading deformation of the slow-roll action is
\beq
\label{equ:Lcs}
{\cal L}_{c_s} = {\cal L}_{\rm s.r.} + \tfrac{1}{2} M_2^4(\varphi) \left[(\partial_\mu \varphi)^2 + 1\right]^2\ .
\eeq
Notice that we added `+1' to the operator in eq.~(\ref{equ:op}) before squaring. This was necessary in order to avoid reintroducing a tadpole for $\pi$.
In other words, without loss of generality, we chose the cancel the tadpole once and for all at lowest order and only add new operators without tadpoles.
Written in terms of the Goldstone $\pi$, we see that the operator in eq.~(\ref{equ:Lcs})
modifies the kinetic term, but {\it not} the gradient term
\beq
{\cal L}_{c_s} = - (\Mp^2 \dot H - 2 M_2^4) \dot \pi^2 + \Mp^2 \dot H (\partial_i \pi)^2 + \cdots
\eeq
This induces a sound speed for the propagation of $\pi$,
\beq
\frac{1}{c_s^2} \equiv 1 - \frac{2 M_2^4}{\Mp^2 \dot H}\ ,
\eeq
with a large value for the parameter $M_2$ corresponding to a small sound speed $c_s \ll 1$.
Substituting (\ref{equ:op}) into (\ref{equ:Lcs}), we see immediately that a small sound speed relates to large interactions~\cite{Cheung}
\beq
{\cal L}_{c_s} = \cdots +   2 M_2^4 \left( \dot \pi^2 - \tfrac{1}{2} \dot \pi (\partial_\mu \pi)^2 + \cdots \right)\ .
\eeq
This non-linearly realized symmetry is particularly clear in the EFT approach.

\vskip 4pt
{\it $P(X)$ theories.} \hskip 4pt
The generalization of the single-derivative Goldstone action to so-called $P(X)$ theories \cite{ArmendarizPicon:1999rj,Garriga:1999vw, Chen:2006nt} is straightforward:
 \beq
 \label{equ:PX}
{\cal L}_{P(X)} = \sum_{n=0}^{\infty} \tfrac{1}{n!} M_n^4(\varphi) \left[ (\partial_\mu \varphi)^2 + 1\right]^n  \ .
\eeq
Here, the operators proportional to $M_n$ start at order $n$ in $\pi$ and end at order $2n$.  Only two independent operators---proportional to $M_2$ and $M_3$---therefore contribute to the cubic Lagrangian for $\pi$.

\vskip 4pt
{\it Higher-derivative theories, ghosts and galileons.} \hskip 4pt
The number of operators increases rapidly if we allow for the possibility of more than one derivative acting on $\varphi$.
Higher derivatives lead to interesting theories such as ghost inflation~\cite{ArkaniHamed:2003uz} and galileon inflation~\cite{Burrage:2010cu}.
We won't treat them in this paper, although our formalism of course applies to all of these cases.

\vskip 4pt
{\it Multi-field generalizations.} \hskip 4pt
Finally, we could consider coupling the adiabatic mode $\varphi$ to additional (isocurvature) fields $\sigma$. To keep these extra fields naturally light either requires additional global symmetries or supersymmetry (with a modest amount of fine-tuning) \cite{Senatore:2010wk}.

\vskip 4pt
{\it Supersymmetry.} \hskip 4pt
In the rest of this section we will extend the effective theory of inflation to include supersymmetry.
In \S\ref{sec:break}, we introduce the basic ingredients of the theory and explain our assumptions.
In \S\ref{sec:action}, we present a systematic treatment of effective actions with global supersymmetry.
We discuss supergravity couplings in \S\ref{sec:sugra}. Finally, in \S\ref{sec:extra}, we comment on models with additional matter fields.

\subsection{Supersymmetry Breaking during Inflation}
\label{sec:break}

It will be crucial for the construction of the SUSY EFT to understand how SUSY is broken during inflation.

\vskip 4pt
{\it Vacuum energy.} \hskip 6pt  
The primary source of spontaneous SUSY breaking during inflation is, of course, the positive vacuum energy associated with the de Sitter background.  Although the vacuum energy is present in any model of inflation, it is only by introducing supersymmetry that it breaks a symmetry.
We characterize this effect through a SUSY breaking spurion
\beq
X = x + \sqrt{2}\theta \psi_x + \theta^2 {\cal F}_X\ ,
\eeq 
with nearly constant F-term\footnote{We do not consider D-term breaking. In fact, mostly D-term breaking has been shown to be incompatible with well-defined current multiplets \cite{Komargodski:2009pc,Dumitrescu:2010ca}. By restricting to $F$-term breaking, we are therefore only ignoring models with comparable D-term and F-term contributions.}, $|{\cal F}_X|^2 \sim 3 \Mp^2 H^2$.
The action for $X$ in the rigid limit is
\beq\label{eqn:lx}
\mathcal{L}_X = \int d^4 \theta\, X^\dag X - \left( \int d^2 \theta\,  \sqrt{3} \Mp H \, X + {\rm h.c.}  \right) \ .
\eeq
As written, this action does not generate a mass for $x$.  One could add higher-order terms in the K\"ahler potential---e.g.~$\delta K = \Lambda^{-2}(X^\dag X)^2$---to stabilize the vev at $x= 0$. In any model with $m_x \gg H$, one can arrive at the same results without changing the action by imposing the constraint $X^2 = 0$ \cite{Komargodski:2009rz}.  In the following, we will assume that the classical vev of $x$ vanishes.

For a given inflationary model, the inflaton may or may not be part of the multiplet $X$.  However, supersymmetry is most useful when it protects the mass of the inflaton to the lowest possible scale, namely the Hubble scale $H$.  The F-term for $X$ breaks SUSY at a much higher scale, making it more difficult to protect the inflaton mass when the inflaton is in the same multiplet (see \cite{Baumann:2010ys,Baumann:2010nu} for future discussion and examples).  We will therefore make the further assumption that the inflaton $\phi$ is in a separate multiplet, 
\beq
\label{equ:inflaton}
\Phi = \phi + \sqrt{2}\theta \psi_\phi + \theta^2 F\ ,
\eeq
where 
\beq
\phi \equiv \tfrac{1}{\sqrt{2}} (\sigma + i \varphi_c)  \ , \quad {\rm with} \qquad \varphi_c^2 \equiv 2 \Mp^2 |\dot H| (t+\pi)^2\ .
\eeq
We note that supersymmetry inevitably adds a new real scalar degree of freedom $\sigma$. The interactions between $\sigma$ and $\pi$ will be of considerable interest for the phenomenology of inflationary fluctuations.

\vskip 4pt
{\it Time-dependent backgrounds.} \hskip 6pt  The feature that distinguishes inflation from pure de Sitter is that time translations are spontaneously broken by the inflationary background.
This necessarily leads to a second source of SUSY breaking.
We can see this directly from the SUSY algebra $\{ Q_{\alpha} , \bar Q_{\dalpha} \} = 2  \sigma^{\mu}_{\alpha \dalpha} P_{\mu}$.  The breaking of time translations means that the generator $P_0$ is broken.  To be consistent, supersymmetry must also be broken.  This can also be understood from the usual argument that a supersymmetric state must have zero energy.  A rolling scalar field generates positive kinetic energy and breaks SUSY even in the absence of potential energy.
Finally, we note that this type of SUSY breaking is reflected in the supersymmetry transformation of the spinor field
\beq
\delta \psi_\phi = i \sqrt{2} \sigma^\mu \bar \xi \partial_\mu \phi + \sqrt{2} \xi F\ .
\eeq
The fermion transforms inhomogeneously and, hence, supersymmetry is spontaneously broken by the time-dependent inflaton vev, $\langle \dot \phi \rangle \ne 0$.  In superspace, the breaking manifests itself as a non-zero $\theta \bar \theta$ component of $\Phi$, i.e.~$\langle \Phi\rangle |_{\theta \bar \theta}   = i \sigma^\mu \langle \partial_\mu \phi \rangle = - \sigma^0 (\Mp^2 |\dot H|)^{1/2} $.  
This type of SUSY breaking is not usually considered in particle physics since it breaks Lorentz invariance. 
We will refer to this type of SUSY breaking as `Lorentz breaking'.  

\subsection{Supersymmetric Actions}
\label{sec:action}

In a rigid theory in flat space, we can always decouple the actions for $X$ and $\Phi$,\footnote{Of course, one cannot decouple $X$ and $t$ completely, otherwise inflation would never end.  However, this coupling may arise indirectly through couplings to fields that are sufficiently massive during inflation that they can be integrated out, leaving the decoupled action to good approximation.  These fields may become light or tachyonic in order to end of inflation without affecting the fluctuations during the $e$-folds that are probed by the CMB and~LSS. }
\beq
{\cal L} = {\cal L}_X + {\cal L}_\Phi\ . \label{equ:decA}
\eeq
What are the constraints on the form of the Lagrangian ${\cal L}_\Phi$?
Having organized the multiplet in terms of $\varphi = t+ \pi$, we have ensured that $\Phi$ is invariant under local time translations.  As a result, any supersymmetric action written as a function of $\Phi$ and $\Phi^\dag$ is time translation invariant, including terms with any number of derivatives.  However, as we saw in \S\ref{sec:adi}, tadpole cancellation imposes additional constraints on the form of the action.  This is where the EFT of inflation differs from the effective action for Goldstone bosons of global symmetries, where the only constraint on the action is that it is invariant under the symmetry.
Finally, we will impose an additional shift symmetry, $\pi \to \pi + const.$, on the action to ensure at least 60 $e$-folds of scale-invariant fluctuations for $\zeta = - H \pi$.
The most general\footnote{ We have actually imposed a slightly stronger condition, namely that this is a symmetry of the K\"ahler potential itself.  Here, we are anticipating that supergravity effects can break symmetries of the action that are not symmetries of the K\"ahler potential.} Lagrangian of $\Phi$ with this symmetry is\footnote{In flat space, there is no need for additional terms to cancel tadpoles for $\pi$.  Because of the shift symmetry, any terms linear in $\pi$ are total derivatives in flat space.  However, we will be most interested in FRW backgrounds, where tadpole cancelation requires that we consistently couple to gravity.  }
\beq
\label{equ:singleAct}
{\cal L}_\Phi = \int d^4 \theta\, K(\Phi + \Phi^\dag, \partial_\mu \Phi, \partial_\mu \Phi^\dag, \cdots) \ .
\eeq
For example, in slow-roll inflation in the decoupling limit, the unique action is
\beq\label{eqn:susysr}
{\cal L}_{\rm s.r.} = \int d^4 \theta\, \tfrac{1}{2} (\Phi + \Phi^{\dagger})^2 + {\cal L}_X\ .
\eeq
More general theories are characterized by higher-derivative corrections to (\ref{eqn:susysr}). For instance, in \S\ref{sec:cs}, we will study the following example
\begin{align}
{\cal L}_{c_s} 
 = \int d^4 \theta\, \tfrac{1}{2} (\G + {\G}^\dag)^2 \Big[ c_1+  \frac{c_2}{\Mp^2 |\dot H|}\, \partial_\mu  \G \partial^\mu {\G}^\dag \Big]  + {\cal L}_X \ ,
\end{align}
corresponding to a supersymmetric theory with non-trivial sound speed.

\subsection{Coupling to Supergravity}
\label{sec:sugra}

To describe the dynamics of inflation, we must couple the theory to gravity.  For a supersymmetric theory, this means coupling to supergravity.  The decoupling between $\Phi$ and $X$ that was possible for the rigid theory in flat space is no longer possible.\footnote{It is amusing to note the similarity between the SUSY EFT of inflation and Goldstini~\cite{Cheung:2010mc}, as both describe two independent SUSY breaking sectors that are coupled only through gravity.  These theories are different in two respects:  First of all, in inflation, the massive scalar $\sigma$ will not, in general, be sufficiently massive that it can be integrated out and will play a critical role in the phenomenology of the EFT.  Second of all, one of the sources of breaking is `Lorentz breaking' rather than F-term breaking.  Even in models in which $\sigma$ is sufficient heavy that it can be integrated out, the Goldstino multiplet takes the form $\Phi_{\rm NL} = \psi \sigma^0 \bar \psi / \Lambda_b^2 + \varphi + \ldots$ , where $\psi$ is the Goldstino of the Lorentz breaking sector and $\Lambda_b$ is the scale where the symmetry is broken.  Unfortunately, since fermions don't freeze out,  the Goldstini are unobservable.} 

\vskip 4pt
{\it Tadpole cancellation.} \hskip 6pt
The decoupling between $\Phi$ and $X$ is broken by the time evolution of the vacuum energy, $\dot H \ne 0$.     In order to reproduce the correct time-dependent vacuum energy, $\Lambda(\varphi) = \Mp^2(3 H^2 (\varphi) + \dot H)$, we include a small superpotential coupling between $\Phi$ and $X$, 
\beq
\mathcal{L}_X \to \int d^4\theta\, X^{\dagger} X -  \left(  \int d^2 \theta \, v(-i \Phi ) X +{\rm h.c.} \right) \ ,
\eeq
where  $|v(\varphi)|^2 = \Mp^2(3 H^2 (\varphi) + \dot H)$.  This is equivalent to including couplings that cancel the tadpoles for $\pi$ at all values of $t$.  
Additional couplings may be required to cancel tadpoles that may arise for the inflaton partner $\sigma$. 

\vskip 4pt
{\it Curvature couplings.} \hskip 6pt
Further couplings between $\Phi$ and $X$ generically arise in supergravity
\beq
\mathcal{L}_{X \Phi} \ \supset\ \alpha\, \mathcal{R} \sigma^2+\cdots + \beta \int d^4 \theta\, \frac{1}{\Mp^2} (\Phi+\Phi^{\dagger})^2 X^{\dagger}X + \cdots \ , \label{equ:sug}
\eeq
where $\alpha$ and $\beta$ are dimensionless coupling constants of order unity.
Notice that only the non-shift-symmetric part of $\Phi$, i.e.~$\sigma \propto \Phi + \Phi^\dag$, receives corrections, while the potential for the shift-symmetric part,  i.e.~$\varphi \propto \Phi - \Phi^\dag$, remains protected.
Although curvature couplings, such as the first term in (\ref{equ:sug}), do not couple the fields directly, Einstein's equations relate the spacetime curvature $\mathcal{R}$ to the vacuum energy generated by $X$.
This indirectly couples $X$ and $\Phi + \Phi^\dag$.  The second term  in (\ref{equ:sug}) represents Planck-suppressed couplings between the two sectors.  Using ${\cal F}_X \sim \sqrt{3} \Mp H$ and $\mathcal{R} = -12 H^2$ (here we are using the sign conventions of~\cite{WB}), we see that both terms in (\ref{equ:sug}) contribute a mass to $\sigma$ of order $H$.\footnote{In the limit, $\Mp \to \infty$, the second class of corrections naively seem to vanish, while the first class remains.  However, for the background to solve Einstein's equations for any finite value of $\Mp$, we must take ${\cal F}_X \to \infty$ in the same limit.  For this reason, both types of terms survive the decoupling limit and contribute terms of order $H^2$.  This non-decoupling of SUSY breaking is necessary.  In de Sitter space, the transformation of the gravitino is required for any action to be supersymmetric and therefore de Sitter space must break supersymmetry explicitly in the decoupling limit. }  

\vskip 4pt
{\it Auxiliary fields in supergravity.} \hskip 6pt
Minimal supergravity has two auxiliary fields, a complex scalar $M$ and a real vector $b_{\mu}$~\cite{WB} (see Appendix~\ref{sec:SUGRA}).
These fields can also couple to the inflaton. When $M$ and $b_{\mu}$ acquire vevs, they can, in principle, affect the inflationary dynamics.
For models of particle physics (e.g.~gravity or anomaly mediation), the contribution to the soft masses from $M$ is extremely important: in this case, a vev of the superpotential $W$ leads to  a universal contribution of the form $\langle M \rangle \sim \langle W\rangle / \Mp^2$. Such a vev, in fact, has to be turned on in order to achieve a small value for the present day cosmological constant, i.e.~$W_0 \sim {\cal F}_X \Mp$ and hence $\langle M\rangle \sim {\cal F}_X / \Mp$.  Therefore, SUSY breaking effects communicated by $M$ are comparable to those from Planck-suppressed mixing between sectors.
In inflation, the situation is very different:  The vacuum energy plays a crucial role and should not be cancelled by the vev of the superpotential.  As a result, one requires that $\langle W\rangle \ll {\cal F}_X \Mp$ and therefore $\langle M\rangle \ll {\cal F}_X/\Mp \sim H$. This makes the contributions from $M$ subdominant relative to the generic curvature couplings from supergravity.  

In fact, under the assumptions stated in \S\ref{sec:break}, we find that the vevs of the auxiliary fields $M$ and $F$ are strictly zero.
The basic reason for this is as follows:  The action for the SUSY effective theory of inflation has an $R$-symmetry.  Under this symmetry, $X$ has $R$-charge 2 and $\Phi$ has $R$-charge 0.  As a result, both $\varphi$ and ${\cal F}_X$ have charge zero and neither vev breaks the $R$-symmetry.  However, both $M$ and $F$ are charged and a vev for either would break the symmetry.  Since $M$ and $F$ are auxiliary fields, any vev should be proportional to an $R$-symmetry breaking parameter.  The only available quantity is the vev of $x$.  However, if $\langle x\rangle=0$, as we assume throughout, then $\langle M \rangle= \langle F \rangle = 0$.

In contrast, there is no general principle that requires $\langle b_{\mu} \rangle = 0$. However, in all of the models that we will discuss in this paper, we will actually  find this to be the case.  More generally, a vev for $b_{\mu}$ breaks diffeomorphism invariance.  Because time diffeomorphisms are broken, $b_0$ can acquire a vev.  A vev for $b_0$ also breaks SUSY and must be related to the ``Lorentz -breaking" of supersymmetry.  Because this is a subdominant contribution to the vacuum energy, $\langle b_0 \rangle \ll H$ during inflation.  

\vskip 4pt
{\it Supergravity for effective theories.} \hskip 6pt
In this section we have merely outlined the contributions to the action that may arise from supergravity.  Determining the precise form of these corrections requires a complete treatment of supergravity, including a careful treatment of higher-derivative contributions to the K\"ahler potential.  We summarize the basic results of such an analysis in Appendix~\ref{sec:SUGRA}. Full details can be found in a companion paper~\cite{Paper1}.

\subsection{Additional Superfields}
\label{sec:extra}

So far, we have introduced supersymmetry with a minimal field content of two chiral fields, $X$ and $\Phi$.  Of course, there is nothing that forbids additional chiral fields, $\Sigma_i$, from appearing in the action.  Any sector that is decoupled from $X$ will be protected from quadratic divergences just like the inflaton.  It is a special feature of SUSY that we do not need to enlarge the symmetry group to protect additional scalar fields.
We wish to treat these additional fields, $\Sigma_i = \tilde \sigma_i+ \sqrt{2} \theta \tilde \psi_i + \theta^2 \tilde F_i$, as part of the inflationary sector.  Therefore, in the rigid limit, we may write the action as\footnote{A subset of these models with ${\cal L}_{\Phi \Sigma} = 0 $ was studied in \cite{Senatore:2010wk}.  For such models to leave an imprint on inflation, the mass of $\tilde \sigma_i$ was required to be $m^2_{\tilde \sigma_i} \lesssim 10^{-2} H^2$ and the fluctuations in $\tilde \sigma_i$ were converted to curvature fluctuations $\zeta$ {\it after} inflation.  For a comprehensive discussion of the construction and signatures of these models we refer the reader to \cite{Senatore:2010wk}.}
\beq
{\cal L} =  {\cal L}_X + {\cal L}_\Phi + {\cal L}_{\Sigma} + {\cal L}_{\Phi \Sigma} \ .
\eeq
Here, we assumed a decoupled action between $X$ and $\Sigma_i$, just as we did for $\Phi$.  We will demand that $\langle\tilde \sigma_i \rangle = 0$ and $\langle \tilde F_i \rangle = 0$ during inflation.  Just as in the case of $\Phi$, coupling to supergravity will induce masses of order $H^2$ for $\tilde \sigma_i$ (unless we impose an extra global symmetry for $\tilde \sigma_i$). 
Making no further assumptions, we write the action as
\beq\label{eqn:multifield}
{\cal L}_\Phi + {\cal L}_{\Sigma} + {\cal L}_{\Phi \Sigma} \ =\ \int d^4 \theta\, K(\Sigma, \Sigma^\dag, \Phi + \Phi^\dag; \partial_\mu \Phi, \partial_\mu \Sigma, \cdots ) + \left( \int d^2 \theta\,  W(\Sigma) + {\rm h.c.} \right)\ .
\eeq
We have dropped the flavor index $_i$, since there is no qualitative differences between one and many $\Sigma$'s. The imaginary part of $\Phi$ is shift-symmetric, as before.

Let us highlight some features that distinguish the action for $\tilde \sigma$---eq.~(\ref{eqn:multifield})---from the action for $\sigma$---eq.~(\ref{equ:singleAct}):
\begin{enumerate}
\item[1)] First, we note that we can have a superpotential for $\Sigma$, while the shift symmetry forbids a superpotential for $\Phi$. This implies we can write SUSY-preserving, non-derivative interactions for $\tilde \sigma$.  For example, the interaction $\lambda | \tilde \sigma|^4$ is obtained from the superpotential $W = \lambda \Sigma^3$.  However, the same interaction for $\sigma$ is only obtained from $\int d^4 \theta\, (\G + \Gd)^6$ or $\int d^4 \theta\, (\G+ \Gd)^4 X^{\dag} X$. In this case, one generates $\lambda \sigma^4$ with $\lambda \propto \int d^4 \theta\, \mathcal{O} \neq 0$, where $\mathcal{O} = \{X^{\dagger} X, (\Phi+ \Phi^{\dagger})^2, \cdots \}$ is an operator whose D-term breaks SUSY.
\item[2)] Finally, there are differences in the form of derivative couplings.  For example, the coupling $\partial_{\mu} \varphi \partial^\mu \sigma$ can only be written by using a SUSY breaking D-term, since $\sigma$ and $\varphi$ appear in the same multiplet.  However, with $\tilde \sigma$ in a different chiral multiplet, the coupling $\partial_{\mu} \varphi \partial^\mu (\tilde \sigma + \tilde \sigma^*)$ is easily achieved using $\int d^4 \theta\,  i (\G+ \Gd) (\Sigma -\Sigma^{\dag}) $.
This observation will be of some significance in our constructions of supersymmetric quasi-single-field inflation in \S\ref{sec:SUSY_QSFI}.
\end{enumerate}

\vskip 6pt
In this section, we have developed a general framework for supersymmetrizing the effective theory of inflation.
In the next two sections, we will apply this formalism to concrete examples: in Section~\ref{sec:cs}, we study naturalness in SUSY theories with small $c_s$, while in Section~\ref{sec:QSFI}, we construct explicit realizations of SUSY QSFI. Both sections are self-contained, so they may be read in any order. In particular, readers whose main interests lie in the observational signatures of SUSY are advised not to let themselves be distracted by the technical details of Section~\ref{sec:cs} and instead skip straight to Section~\ref{sec:QSFI}.

\section{Naturalness and Small Sound Speed}
\label{sec:cs}

We now turn to our first application of the SUSY EFT of inflation: supersymmetric theories with small sound speed and large derivative interactions.
As we saw in \S\ref{sec:adi}, at cubic order in fluctuations, there are only two single-derivative operators deforming the slow-roll action
\beq
{\cal L} \ =\  \Mp^2 \dot H (\partial_\mu \varphi)^2 + \tfrac{1}{2} M_2^4 \left[(\partial_\mu \varphi)^2 + 1\right]^2 + \tfrac{1}{3!} M_3^4 \left[(\partial_\mu \varphi)^2 + 1\right]^3 + \cdots \ , \label{equ:LL}
\eeq
where
\beq
\frac{1}{c_s^2} = 1 -\frac{2 M_2^4}{\Mp^2 \dot H} \ .
\eeq
In this section, we will realize these theories (and their generalizations to $P(X)$ theories) in supersymmetry (see also \cite{Justin}).  
In \S\ref{sec:SUSYPX}, we derive the component Lagrangians for those theories and show that 
higher-derivative couplings lead to a parametrically enhanced mass for the SUSY partner of the inflaton, 
\beq
m_\sigma^2 \sim \frac{H^2}{c_s^2} \gg H^2\ .
\eeq
We then use the SUSY EFT to revisit constraints on the natural sizes of the parameters $M_2$ and $M_3$.
Specifically, in \S\ref{sec:weak}, we ask whether the hierarchy $M_2^4 \gg \Mp^2 |\dot H|$ (and hence $c_s \ll 1$) is techincally natural. To answer the question, we construct a weakly-coupled UV-completion of theories with non-trivial sound speed. We then proceed to show that small $c_s$ is unnatural in the non-supersymmetric theory, but becomes natural when the theory is made supersymmetric.
Finally, in \S\ref{sec:ortho}, we discuss the natural size of the parameter $M_3$ relative to the size of $M_2$. We will comment on the implications of our results for the naturalness of orthogonal type non-Gaussianity, which relies on a cancellation between the two operators in (\ref{equ:LL}).

\subsection{Supersymmetric $P(X)$ Theories}
\label{sec:SUSYPX}

{\it Supersymmetric sound speed.} \hskip 6pt
We begin with the Lagrangian for a theory with small $c_s$,
\beq
{\cal L}_{c_s}  = -\Mp^2(3H^2+\dot H) + \Mp^2 \dot H (\partial_\mu \varphi)^2 + \tfrac{1}{2} M_2^4 \left[(\partial_\mu \varphi)^2 + 1 \right]^2 \ .
\eeq
In order to supersymmetrize the theory it proves useful to slightly reorganize the terms
\beq
{\cal L}_{c_s}  = \Big(\underbrace{-\Mp^2(3H^2+\dot H)+ \tfrac{1}{2}M_2^4 }_{m_0^4}\Big) + \Big(\underbrace{\Mp^2 \dot H + \tfrac{1}{2}M_2^4 }_{m_1^4} \Big)(\partial_\mu \varphi)^2 +\tfrac{1}{2}  \underbrace{ M_2^4}_{m_2^4}\, (\partial_\mu \varphi)^2 \left[(\partial_\mu \varphi)^2 + 1 \right] \ . \label{equ:LLLs}
\eeq
The reason for writing the Lagrangian in this way is that $(\partial_\mu \varphi)^2$ and $(\partial_\mu \varphi)^2 \left[(\partial_\mu \varphi)^2+1\right]$ each correspond to individual operators in SUSY, while $\left[(\partial_\mu \varphi)^2+1\right]^2$ does not.
The Lagrangian (\ref{equ:LLLs}) arises from the following higher-derivative K\"ahler potential\footnote{Of course, there is no unique supersymmetric generalization of these theories. There are various actions that differ in the couplings of fermions and additional scalars, but that reduce to the same theories when those extra fields have vanishing vevs.
The different ways of embedding small $c_s$-theories in supersymmetric actions were explored in~\cite{Justin}. In particular, they studied an interesting alternative operator that also generates a small sound speed, 
\beq
K_{c_s}  = \tfrac{1}{2} \hat c_1 (\G+ \Gd)^2 +  \tfrac{1}{16} \hat c_2\, D^{\alpha} \G D_{\alpha} \G \bar D_{\dalpha} \Gd \bar D^{\dalpha} \Gd \ .  \nonumber
\eeq
As usual, the first term is only shift symmetric for the imaginary part of $\G$.  However, the second term in invariant under shifts of both the real and imaginary parts. To see this note that $D_{\alpha}( \G \pm \Gd) = D_{\alpha} \G$, using the fact that $\G$ is chiral.} 
\begin{align}
\label{equ:example}
K_{c_s} 
\ =\ \tfrac{1}{2} (\G + {\G}^\dag)^2 \Big[ c_1+  \frac{c_2}{\Mp^2 |\dot H|}\, \partial_\mu  \G \partial^\mu {\G}^\dag \Big]  \ , 
\end{align}
where 
\beq
c_1 \equiv - \tfrac{m_1^4}{\Mp^2 |\dot H|} =  1 + \tfrac{1}{2} \big( 1- \tfrac{1}{c_s^2}\big) \qquad {\rm and} \qquad c_2 \equiv - \tfrac{m_2^4}{\Mp^2 |\dot H|}  = \tfrac{1}{4} \big( 1 - \tfrac{1}{c_s^2}\big)\ .
\eeq
 The superpotential is the same as in slow-roll inflation,
\begin{align}
W_{c_s} &=  - v(i \hat \Phi) X\ , \label{equ:Wcs}
\end{align}
but now $|v(\varphi)|^2  \equiv - m_0^4 $.

\vskip 6pt
{\it Curvature couplings.} \hskip 6pt
In our companion paper~\cite{Paper1}, we derive the complete supergravity action for eq.~(\ref{equ:example}), including all couplings to the auxiliary field $F$, $M$, and $b_\mu$.
However, in \S\ref{sec:sugra}, we argued that the vevs for the auxiliary fields are either suppressed or strictly zero.
We therefore only show the result for the dominant curvature couplings~\cite{Paper1}:
\beq
{\cal L}_{c_s} \ =\  - |v|^2 + c_1 {\cal L}_1 + \frac{c_2}{ \Mp^2 |\dot H|}\, {\cal L}_2\ ,
\eeq
where 
\begin{align}
{\cal L}_1 &= -  |\partial_\mu \phi|^2 + \Big(\underline{\underline{\tfrac{1}{6} {\cal R} - \tfrac{1}{3\Mp^2} |{\cal F}_X|^2}} \Big) \sigma^2\ , \label{equ:L1xx} \\
{\cal L}_2 &= -  \left( |\partial_\mu \phi|^2 \right)^2 - 2 \partial_\mu \sigma \partial_\nu \sigma   \partial^\mu \phi \partial^\nu \phid \nonumber  \\
&\hspace{0.4cm}+ \sigma^2 \Big\{  \Big( \, \underline{\underline{\tfrac{1}{6} {\cal R} -  \tfrac{1}{3\Mp^2} |{\cal F}_X|^2}} \, \Big) \big[ (\partial_\mu \varphi_c)^2 + (\partial_\mu \sigma)^2\big]  - \tfrac{1}{2} \underline{\underline{{\cal R}_{\mu \nu}}}  \big[ \partial^\mu \varphi_c \partial^\nu \varphi_c + \partial^\mu \sigma \partial^\nu \sigma\big] \nonumber \\
&\hspace{1.7cm}  + \tfrac{1}{2} \underline{\underline{\nabla_\mu \nabla_\nu}} \big[ \partial^\mu \varphi_c \partial^\nu \varphi_c + \partial^\mu \sigma \partial^\nu \sigma\big] +  \partial^\mu \phi \partial_\mu \Box \phid + \tfrac{1}{2}\Box|\partial_\mu \phi|^2 \Big\} \ . \label{equ:L2xx}
\end{align}
Using $\bar \varphi_c = (2\Mp^2 |\dot H|)^{1/2} t$ and ${\cal R}_{\mu \nu} = -3H^2 g_{\mu \nu}$ (de Sitter), we find that only four terms in (\ref{equ:L1xx}) and (\ref{equ:L2xx})---see underlined---contribute a mass to the field $\sigma$, 
\beq
{\cal L}_{c_s} = \cdots +\left[ \, c_1\left(\tfrac{1}{6} {\cal R} - H^2 \right) + c_2 \left( \tfrac{1}{12}  {\cal R} - 2 H^2 \right) (\partial_\mu t)^2 + c_2 \nabla_\mu \nabla_\nu (\partial^\mu t \partial^\nu t) \right] \sigma^2  + \cdots \ .
\eeq
With ${\cal R} = -12 H^2$ and $\nabla_\mu \nabla_\nu (\partial^\mu t \partial^\nu t) = 9H^2$, we find
\beq
m_\sigma^2 = 6H^2 c_1 -12 H^2 c_2 \simeq 0 \times \frac{H^2}{c_s^2} + 6 H^2 \ . \label{equ:MS2}
\eeq
The cancellation of the leading $H^2 c_s^{-2}$ terms that we find here is completely accidental and not protected by any symmetry.
In particular, as we explain in Appendix~\ref{sec:SUGRA}, in principle, there are a host of additional Planck-suppressed couplings between $\Phi$ and $X$ in the K\"ahler potential, cf.~eq.~(\ref{equ:sug}). These terms are model-dependent, but typically they lead to large contributions to the mass, 
\beq
m_\sigma^2 \sim
\frac{H^2}{c_s^{2}}\ .
\eeq
We therefore conclude that generically the mass of the partner of the inflaton is enhanced in the limit of small $c_s$. This is important since it implies that $\sigma$ won't receive quantum mechanical fluctuations during inflation.

\vskip 8pt
\small
\hrule
\vskip 1pt
\hrule
\vskip 6pt
\small
\noindent
{\it SUSY $P(X)$.}  \hskip 6pt
The above is easily generalized to $P(X)$ theories.
Again, it is useful to first re-write the non-SUSY effective action
\begin{align}
\label{equ:csX}
{\cal L}_{P(X)} &= m_0^4(\varphi) + m_1^4 (\partial_\mu \varphi)^2 + \sum_{n=2}^\infty \tfrac{1}{n!} m_n^4 (\partial_\mu \varphi)^2 \left[ (\partial_\mu \varphi)^2 + 1\right]^{n-1} \ .
\end{align}
 This is nothing deep, but just a simple reshuffling of the operators we are familiar with from~\S\ref{sec:adi}:
\begin{align}
{\cal L}_{P(X)}  &= \underbrace{\left( m_0^4 - \tfrac{1}{2} m_2^4 \right)}_{-\Mp^2(3H^2 +\dot H)} +  \underbrace{\left( m_1^4 - \tfrac{1}{2} m_2^4 \right)}_{\Mp^2 \dot H} (\partial_\mu \varphi)^2 + \sum_{n=2}^\infty \tfrac{1}{n!} \underbrace{\left( m_n^4 - \tfrac{m_{n+1}^4}{n+1}  \right)}_{M_n^4} [(\partial_\mu \varphi)^2 + 1]^n\ .
\end{align}
Writing the effective theory in the form of eq.~(\ref{equ:csX}) has the advantage that it is easy to supersymmetrize.
A possible form for the K\"ahler potential is
\begin{align}
K_{P(X)} 
\ =\ - \tfrac{1}{2}(\hat \G + {\hat \G}^\dag)^2 \left[  m_1^4  + \sum_{n=2}^\infty \tfrac{1}{n!} m_n^4 \left[\partial_\mu \hat \G \partial^\mu {\hat \G}^\dag + 1 \right]^{n-1} \right]\ ,
\end{align}
where $\hat \G \equiv (\Mp^2 |\dot H|)^{-1/2} \G$. The superpotential is the same as in eq.~(\ref{equ:Wcs}).
\vskip 6pt
\hrule
\vskip 1pt
\hrule
\vskip 6pt
\normalsize 
\vspace{0.5cm}

\subsection{Naturalness of Small Sound Speed}
\label{sec:weak}

{\it Is the hierarchy $M_2^4 \gg \Mp^2 |\dot H|$---and hence $c_s \ll 1$---natural?}
Before we address this question, we digress briefly to review our prior work on the subject~\cite{Baumann:2011su}.

\vskip 4pt 
{\it Strong coupling and new physics.} \hskip 6pt 
It is well-known \cite{Cheung, Leblond:2008gg, Baumann:2011su} that theories with small sound speed become strongly coupled at energies not too far above the Hubble scale.
Of course, CMB observations probe modes at the Hubble scale, so, in principle, there is no problem with the theory becoming strongly coupled at higher energies.
Nevertheless, it is interesting to take strong coupling as an indication for `new physics' entering before the would-be strong coupling scale.\footnote{The same philosophy can be applied to the Higgless Standard Model. A theory of massive $W$ and $Z$ bosons becomes strongly coupled around 1 TeV, unless some new physics enters before. In that case, we know that a light Higgs keeps the theory weakly coupled and restores perturbative unitarity.} 
The new physics could manifest itself as the appearance of new massive degrees of freedom or as a change in the physical description of the existing degrees of freedom.
In \cite{Baumann:2011su} we applied this logic to inflationary models with small sound speed.
We showed that the theory can stay weakly coupled if the dispersion relation of the inflaton field changes before the would-be strong coupling scale $\Lambda_\star$. We presented a specific two-field theory (see also \cite{Tolley:2009fg,Cremonini:2010ua,Achucarro:2010da}) that allows for large non-Gaussianities at Hubble without strong coupling above Hubble.
However, weak coupling came at a price: the theory only leads to a small sound speed for unnatural mass parameters and couplings.\footnote{Again, this has a parallel in the Standard Model: a light Higgs is unnatural unless something---such as SUSY---controls quantum corrections to the Higgs mass.}
In this section we show that the theory, in fact, becomes natural in the presence of SUSY.

\vskip 4pt
{\it A weakly-coupled UV-completion.} \hskip 6pt
 Here, we present the basic equations of our weakly-coupled UV-completions of theories with small $c_s$.
 For further details we refer the reader to our previous publication~\cite{Baumann:2011su}.
 
Consider adding a second massive field $\sigma$ to the slow-roll Lagrangian
\begin{align}
\mathcal{L}_0 &=  - \tfrac{1}{2} (\partial_\mu \pi_c)^2 - \tfrac{1}{2} (\partial_\mu \sigma)^2 - \tfrac{1}{2} m_\sigma^2 \sigma^2 \ ,  \label{equ:L0}
\end{align}
where $\pi_c^2 \equiv 2\Mp^2 |\dot H| \pi^2$. In order for $\sigma$ to affect the dynamics of $\pi_c$, we couple the two fields through the following interaction
\beq
{\cal L}_{\rm mix} = - m^3 \left[(\partial_\mu \varphi)^2 + 1 \right] \sigma \ \to\ \rho \Big( \dot \pi_c - \frac{1}{2} \frac{(\partial_\mu \pi_c)^2}{(2\Mp^2 |\dot H|)^{1/2}} \Big) \sigma\ . \label{equ:MIX}
\eeq
At high energies, $\omega > \rho$, this describes two decoupled fields $\pi_c$ and $\sigma$, with a small perturbative mixing $\rho \dot \pi_c \sigma$\,:
\beq
\includegraphicsbox[scale=1.0]{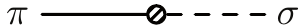}
\eeq
  \begin{wrapfigure}{r}{.22\textwidth}
\vspace{-0.5cm}
\begin{center}
\includegraphics[width=0.125\textwidth]{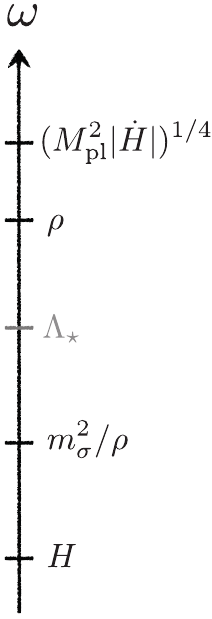}
\end{center}
\vspace{-0.2cm}
\end{wrapfigure}  
At lower energies, $\omega < \rho$, the mixing term dominates the dynamics. The theory becomes a non-relativistic {\it single-field} theory (if this is counterintuitive please see \cite{Baumann:2011su} for a detailed discussion of this feature of the theory). In this regime, the disperision relation for $\pi$ is non-linear, $\omega = k^2 / \rho$. 
For $\rho^4 < M_{\rm pl}^2 |\dot H|$, the theory is weakly coupled at all energies up to the symmetry breaking scale $ \Mp^2 |\dot H|$.
At even lower energies, $\omega < m_\sigma^2/\rho$, the mass term of $\sigma$ becomes relevant and the theory develops a linear dispersion, $\omega =c_s k$, with sound speed given by
\beq
c_s \simeq \frac{m_\sigma}{\rho}\ .
\eeq
We see that a small sound speed requires $m_\sigma \ll \rho$.
Is this hierarchy technically natural?

\vskip 4pt
{\it Without SUSY.} \hskip 6pt The symmetries of the EFT relate the mixing term $\rho \dot \pi \sigma$ to a cubic interaction $(\partial_\mu \pi_c)^2 \sigma$, cf.~eq.~(\ref{equ:MIX}). This interaction induces radiative corrections to the mass parameter $m_\sigma$ and the kinetic mixing parameter $\rho$. At one loop, we find:
\vspace{-0.2cm}
\begin{align}
\delta m_\sigma^2 \ =\ \sigma \includegraphicsbox[scale=1.0]{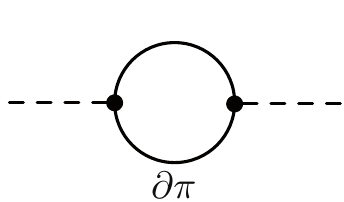}\ \sigma &\ \sim\ \frac{\rho^2}{\Mp^2 |\dot H|}\, \Lambda_{\rm uv}^4 \ \sim\ \rho^2\ , \label{equ:dm} \\
\delta \rho \ =\ \dot \pi\, \includegraphicsbox[scale=1.0]{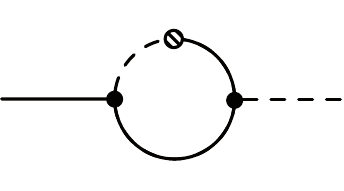}\ \sigma &\ \sim \ \frac{\rho^3}{\Mp^2 |\dot H|}\, \Lambda_{\rm uv}^2 \ \sim \ \frac{\rho^2}{(\Mp^2 |\dot H|)^{1/2}} \, \rho\ . \label{equ:rhoL}
\end{align}
To estimate the size of the loops we have set the UV-cutoff equal to the symmetry breaking scale, $\Lambda_{\rm uv}^4 \sim \Mp^2 |\dot H|$. We see that the weak coupling requirement, $\rho^4 < \Mp^2 |\dot H|$, is natural since the loop correction in (\ref{equ:rhoL}) is suppressed by the ratio $\frac{\rho^2}{(\Mp^2 |\dot H|)^{1/2}} < 1$. However, the hierarchy $m_\sigma < \rho$ and hence $c_s < 1$ is {\it not} natural. Instead, the loop correction to the mass in eq.~(\ref{equ:dm}) implies that the natural value of the sound speed is unity, 
\beq
\delta c_s^2 \simeq \frac{\delta m_\sigma^2}{\rho^2} \sim 1\ .
\eeq

\vskip 4pt
{\it With SUSY.} \hskip 6pt
A supersymmetric version of the theory described by eqs.~(\ref{equ:L0}) and (\ref{equ:MIX}) is
\beq
\label{equ:WeakSusy}
\mathcal{L} = \int d^4 \theta \, \left[ \tfrac{1}{2} (\G + \Gd)^2 + \frac{1}{\Lambda} (\G + \Gd)^3  \right]\ . 
\eeq
The second term in (\ref{equ:WeakSusy}) generates a quadratic kinetic mixing term with $\rho = \frac{(\Mp^2 |\dot H|)^{1/2}}{\Lambda}$. 
Importantly, this breaks supersymmetry at the scale~$\rho$. 
Notice that this SUSY breaking is associated with the time dependence of the background $\bar \varphi = t$ and not with the constant vacuum energy.
Nevertheless, for $\rho^4 < \Mp^2 |\dot H|$, SUSY still helps to regularize the loop correction to the mass. 
We can see this in two different ways:
\begin{enumerate}
\item[1)] Only loops that include the SUSY breaking will contribute to the mass of $\sigma$. The loop that led to eq.~(\ref{equ:dm}) does {\it not} include SUSY breaking since the operator $(\partial_\mu \pi_c)^2 \sigma$ in (\ref{equ:MIX}) is supersymmetric (it isn't proportional to the time-dependent background vev). SUSY will therefore enforce that a fermion loop cancels the boson loop leading to (\ref{equ:dm}). Hence, we only get a contribution to the mass term if the SUSY breaking operator $\dot \pi_c \sigma = (\partial_0 t) \dot \pi_c \sigma$ is included in the loop.
Such a loop then leads to 
\beq
\delta m_\sigma^2 \ =\ \sigma \includegraphicsbox[scale=1.0]{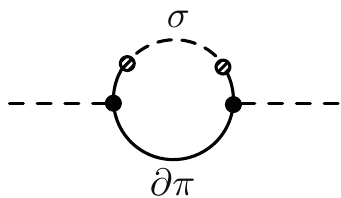}\ \sigma \ \sim\ \frac{\rho^4}{\Mp^2 |\dot H|}\, \Lambda_{\rm uv}^2 \sim \frac{\rho^2}{(\Mp^2 |\dot H|)^{1/2}} \, \rho^2\ . \label{equ:loop1}
\eeq
We see that in the supersymmetric theory the mass of $\sigma$ only has a quadratic divergence, 
while the non-supersymmetric theory had a quartic divergence. This results in the maximal radiative correction to the mass of $\sigma$ now being suppressed by the small ratio $\frac{\rho^2}{(\Mp^2 |\dot H|)^{1/2}} < 1$.
The hierarchy $m_\sigma < \rho$ is therefore natural and protected by SUSY.\footnote{To complete the proof, we should still show that there is no large contribution from modes with energies below the SUSY breaking scale $\rho$. In this regime, eq.~(\ref{equ:dm}) still applies, but the UV-cutoff is now $\Lambda_{\rm uv} \sim \rho$. From modes below the SUSY breaking scale $\rho$, we therefore get the following loop contribution to the mass of $\sigma$,
\beq
\label{equ:mm1}
\delta m_\sigma^2 \ \sim \ \frac{\rho^4}{\Mp^2 |\dot H|} \, \rho^2\ . \nonumber
\eeq
This is smaller than the contribution in (\ref{equ:loop1}) and can therefore be ignored.}
This implies that a small sound speed is technically natural
\beq
\delta c_s^2 \simeq  \frac{\rho^2}{(\Mp^2 |\dot H|)^{1/2}} \ll 1\ .
\eeq
\item[2)] We may arrive at the same result working directly in superspace by treating $ \frac{1}{\Lambda}\int d^4 \theta \,
(\Phi + \Phi^{\dagger})^3$ as a perturbation.  Loops are then computed by using the superspace propagator  to contract vertices.
This leads to the following one-loop correction to eq.~(\ref{equ:WeakSusy}), 
\beq
\label{equ:dL}
\delta {\cal L}_{\rm loop} \ =\ \int d^4 \theta \left[  \frac{\Lambda_{\rm uv}^2}{\Lambda^n} (\G + \Gd)^n +  \frac{1}{\Lambda^n} (\G + \Gd)^{n-2} D^2 \bar D^2  (\G + \Gd)^2 \right] \ +\ \cdots\ ,
\eeq
where the dots indicate terms that are UV-finite.  The appearance of $D^2 \bar D^2$ is the result of acting with the second $\int d^4\theta$ on the external lines, rather than on the propagators.
Eq.~(\ref{equ:dL}) implies the following contribution to the mass of $\sigma$,
\begin{align}
\delta m_\sigma^2 &\ \sim \  \frac{\Lambda_{\rm uv}^2}{\Lambda^4} \Mp^2 |\dot H| \sim \frac{\rho^2}{(\Mp^2 |\dot H|)^{1/2}} \,  \rho^2 \ . 
\end{align}
We again conclude that the hierarchy $m_\sigma < \rho$ is natural and protected by SUSY.
\end{enumerate}

\vskip 4pt
{\it Conclusion.} \hskip 6pt
We have shown that the weakly-coupled UV-completion of $c_s \ll 1$~\cite{Baumann:2011su} is unnatural without SUSY, but becomes natural with SUSY.

\subsection{Naturalness of Orthogonal Non-Gaussianity}
\label{sec:ortho}

{\it What is the natural size of $M_3$?}

\vskip 4pt
We have seen that the cubic Lagrangian for $P(X)$ theories is characterized by two independent operators
\beq
{\cal L} \ = \ \Mp^2 \dot H (\partial_\mu \varphi)^2 +  \tfrac{1}{2} M_2^4 \left[(\partial_\mu \varphi)^2 + 1\right]^2 + \tfrac{1}{3!} M_3^4 \left[(\partial_\mu \varphi)^2 + 1\right]^3\ .
\eeq
In the previous section, we discussed the natural value of $M_2$. We now turn our sights on $M_3$.

\vskip 4pt
In \cite{Senatore:2009gt}, it was argued in the context of the {\it strongly-coupled} UV-completions, that the natural value of the parameter is
\beq
M_3^4 \sim \frac{M_2^4}{c_s^2} \sim \frac{\Mp^2 |\dot H|}{c_s^4} \ .
\eeq
 This conclusion is of observational relevance.
 It implies that the interactions $M_2^4 \dot \pi (\partial_i \pi)^2$ and $M_3^4 \dot \pi^3$ are naturally of similar size and can therefore be cancelled against each other to produce a new bispectrum shape that is orthogonal to the equilateral shape for an order one fraction of the natural parameter space.
In the previous section, we found a natural, {\it weakly-coupled} UV-completion of small~$c_s$ in the context of supersymmetry.  We will now use this theory to determine the natural values of $M_3$.
In order to make contact with \cite{Senatore:2009gt}, we will determine $M_3^4$ for fixed $c_s$ and $\Mp^2 \dot H$, i.e.~we will vary the parameter $\rho$, while simultaneously adjusting $m_{\sigma} = c_s \rho$, such that $c_s$ remains fixed.
As we discussed above, for $\omega < c_s m_\sigma$, the theory describes a single degree of freedom, with $\sigma$ playing the role of the conjugate momentum of $\pi$,
\beq
\sigma \simeq (\Mp^2 |\dot H|)^{1/2} \frac{\rho}{m_{\sigma}^2}\, \dot \pi \ .
\eeq
The low-energy contribution to the operator $M_3^4 \dot \pi^3$ therefore arises predominantly from the coupling $\mu \sigma^3$---i.e.
\beq\label{eqn:Mdef}
M_3^4 \simeq  \frac{(\Mp^2 |\dot H|)^{3/2}}{c_s^6} \frac{\mu}{\rho^3} \ .
\eeq
Here, $\Mp^2 |\dot H|$ and $c_s$ are fixed, so the natural values of $M_3$ are determined by the natural ranges of $\mu$ and $\rho$\,:
\begin{itemize}
\item A lower bound on $\rho$ arises from the requirement that we have a linear dispersion, $\omega = c_sk$, at horizon crossing.  This means that $c_s m_\sigma = c_s^2 \rho > H$.  An upper bound on $\rho$ arises from the lower bound on $m_\sigma$, associated with the minimal loop contribution in eq.~(\ref{equ:loop1}): namely $m_\sigma^2  \gtrsim \delta m_\sigma^2 \sim \rho^2 (\Mp^2 |\dot H|)^{-1/2} \, \rho^2$ leads to $\rho  \lesssim c_s (\Mp^2 |\dot H|)^{1/4}$.
Therefore, we find
\beq
c_s^{-2} H \ \lesssim\ \rho \ \lesssim \ c_s (\Mp^2 |\dot H|)^{1/4} \ .
\eeq

\item The natural range of $\mu$ can be determined by simple loop estimates, as before.  The lower bound is determined from the loop contribution to  $\mu$ arising from the interaction $\rho(\Mp^2 |\dot H|)^{-1/2} (\partial_\mu \pi_c)^2 \sigma$.
As before, the coupling only gets renormalized if we insert the SUSY breaking operator $\rho \dot \pi_c \sigma$ inside the loop
\beq
\delta \mu \ =\  \includegraphicsbox[scale=1.0]{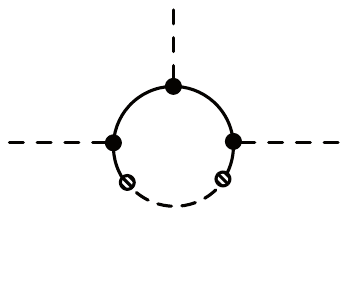} \ \sim\ \frac{\rho^5}{\Mp^2 |\dot H|}\ .
\vspace{-0.5cm}
\eeq
Naturalness of the coupling then requires $\mu \gtrsim \delta \mu \sim \rho^5(\Mp^2 |\dot H|)^{-1}$. 
An upper bound on $\mu$ follows from the renormalization of the mass of $\sigma$, i.e.~$\delta m_\sigma^2 \sim \mu^2$.  Because we have fixed $m_{\sigma}$ in terms of $c_s$ and $\rho$, we must have $\mu \leq m_\sigma$.  
Therefore, we find,
\beq
 \frac{\rho^5}{\Mp^2 |\dot H|} \ \lesssim\ \mu \ \lesssim m_\sigma\ .
\eeq
\end{itemize}
Combining the ranges of $\mu$ and $\rho$ with (\ref{eqn:Mdef}), we find that the natural range of $M_3^4$ is
\beq
\frac{\Mp^2 |\dot H|}{c_s^4}\ \lesssim \ M_3^4 \ \lesssim \ \frac{c_s^{1/2}}{\Delta_\zeta}\, \frac{\Mp^2 |\dot H| }{c_s^4}\ ,
\eeq
where $\Delta_\zeta \equiv H^2/(4\Mp |\dot H|)^{1/2} \sim 10^{-5}$.
We see that we can naturally make the value of $M_3$ larger than the natural value in the strongly-coupled theory \cite{Senatore:2009gt}, but not smaller.

\vskip 4pt
{\it Conclusion.} \hskip 6pt  We have shown that the natural range of $M_3$ is larger in a weakly-coupled SUSY completion than in strongly-coupled models.  Interestingly, the lower bound on $M_3$ is identical in both cases, $M_3^4 \sim \Mp^2 |\dot H| c_s^{-4}$, but the upper bound is now higher.  The orthogonal shape was shown in \cite{Senatore:2009gt} to arise for an order one range of parameters in the strongly-coupled theory.  The fraction of the natural parameter space that gives orthogonal shape is likely to decrease in the weakly-coupled SUSY model.

\section{Supersymmetric Quasi-Single-Field Inflation}
\label{sec:QSFI}

One of the most exciting applications of the supersymmetric effective theory of inflation is to the {quasi-single-field inflation} (QSFI) models of Chen and Wang~\cite{Chen:2009zp}.
These models involve extra fields besides the inflaton field, with masses close to the Hubble scale.
This feature makes supersymmetry a natural arena for QSFI.
In this section, we present a supersymmetric completion of the model of \cite{Chen:2009zp}. Moreover, we use the effective theory approach to propose some obvious generalizations of the original model. We then determine which of these new possibilities can naturally arise in SUSY.
As advertised in the Introduction, QSFI allows for large non-Gaussianities with a unique signature in the squeezed limit. We will show that this signature is robust and not sensitive to details.

\subsection{Quasi-Single-Field Inflation}
\label{sec:QSFIreview}

The basic idea behind QSFI is very intuitive. In this section, we will explain the mechanism and its key predictions in simple physical terms. We then provide a systematic classification of all possible variations of QSFI and show which of them are naturally realized in SUSY.

\subsubsection*{Basic Mechanism and Perturbative Predictions}

 \begin{wrapfigure}{r}{.21\textwidth}
\vspace{-0.4cm}
\begin{center}
\includegraphics[width=0.18\textwidth]{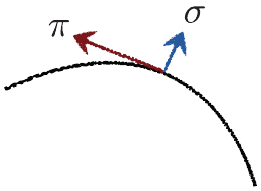}
\end{center}
\vspace{-0.6cm}
\end{wrapfigure}  
\indent
{\it Setup.}  \hskip 6pt QSFI couples a massless Goldstone mode, ${\cal L}_\pi = - \tfrac{1}{2}(\partial_\mu \pi_c)^2$, to a massive isocurvaton mode, ${\cal L}_\sigma = - \tfrac{1}{2} (\partial_\mu \sigma)^2 - \tfrac{1}{2}m_\sigma^2 \sigma^2$, through a quadratic mixing term ${\cal L}_{\rm mix}$, which \cite{Chen:2009zp} chose as ${\cal L}_{\rm mix} =  \rho \dot \pi_c \sigma$. Incidentally, this is the same mixing term that featured prominently in our weakly-coupled UV-completion of theories with small sound speed, cf.~\S\ref{sec:weak}. This coupling arises if the background fields follow a curved trajectory with constant radius of curvature and constant angular velocity. The mixing converts fluctuations in $\sigma$ into fluctuations in $\pi_c$ and hence $\zeta$.
If the mass of the second field is at or below the Hubble scale, $m_\sigma \lesssim \tfrac{3}{2} H$, then quantum fluctuations in $\sigma$ can contribute significantly to the final curvature perturbation.
Moreover, interactions in the $\sigma$-sector are much less constrained than interactions in the inflaton direction.
A large interaction term  ${\cal L}_{\rm int}$ in the $\sigma$-Lagrangian can then be an important source of non-Gaussianity. The specific example ${\cal L}_{\rm int} = -\mu \sigma^3 - \lambda \sigma^4$ was explored in~\cite{Chen:2009zp}.

\vskip 4pt
{\it Power spectrum.}  \hskip 6pt For $\rho < H$, we can treat the mixing term as a perturbative correction, while the leading order dynamics is determined by
\beq
{\cal L}_0 = - \tfrac{1}{2} (\partial_\mu \pi_c)^2  - \tfrac{1}{2} (\partial_\mu \sigma)^2 - \tfrac{1}{2} m_\sigma^2 \sigma^2 \ .
\eeq
The quadratic mixing is then captured by the following transfer vertex
\beq
{\cal L}_{\rm mix} = \rho \dot \pi_c \sigma  \qquad \Leftrightarrow \qquad \includegraphicsbox[scale=1.0]{QSFI_mixing2b} \ .
\eeq
The leading perturbative correction to the power spectrum of $\zeta$ is 
\beq\label{eqn:qsfipower}
\includegraphicsbox[scale=1.0]{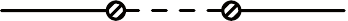} \qquad \Leftrightarrow \qquad  \Delta_\zeta^2 = \frac{1}{4} \frac{H^4}{ \Mp^2 |\dot H|} \left[ 1 + c(\nu) \left( \frac{\rho}{H} \right)^2 \right]\ ,
\eeq
where the precise form of the function $c(\nu)$ can be found in \cite{Chen:2009zp}. Since the correction is scale invariant, it only leads to an unobservable shift in the overall normalization of the power spectrum.

\vskip 10pt
\small
\hrule
\vskip 1pt
\hrule
\vskip 6pt
\small
{\it Scale invariance.}  \hskip 6pt It is worth digressing briefly to explain why QSFI does not introduce large violations of scale invariance.  Since the mixing term couples the Goldstone boson $\pi$ to a massive field $\sigma$, one might suspect that the Goldstone boson effectively becomes massive, leading to a violation of scale invariance in the power spectrum of order $\rho^2 / H^2$. Moreover, one may worry that the violation of scale invariance in the $\sigma$-sector, of order $m_\sigma^2/H^2$, gets communicated to the $\pi$-sector. However, the result of the explicit calculation in (\ref{eqn:qsfipower}) shows no violation of scale invariance.  What happened?

Exact scale invariance of the power spectrum is the result of an exact global symmetry under which $t \to t+ c$ and $k \to k e^{H c}$ (in de Sitter space).  The transformation on $t$ can be undone by a time diffeomorphism, after which one is left with an equivalent global symmetry $\pi \to \pi + c$.  If this symmetry is unbroken, then every mode experiences the same history and the power spectrum of $\pi$ is scale invariant.\footnote{In most models this global time translation symmetry is only approximate, being weakly broken by the time evolution of the Hubble scale, $\dot H \ne 0$.}
In QSFI, the action for $\pi$ and $\sigma$ posses such a symmetry and hence there is no reason to expect any violation of scale invariance.  

This explanation, while true, does not fully address the concern regarding massive fields.  If the  isocurvaton fluctuations, $\sigma$, were converted into curvature perturbations, $\zeta = - H \pi$, {\it after} inflation ends, then we would indeed find violations of scale invariance.  This is because the symmetry $t \to t+ c$ is also broken in order to end inflation.  The evolution of a mode that is frozen outside the horizon is insensitive to the end of inflation and no violation of scale invariance appears.  However, for modes that do not freeze out (i.e.~massive modes), the conversion to curvature perturbations at the end of inflation picks out a scale, since the amplitudes of the different modes are measured at a specific time.
For QSFI, the end of inflation does not induce a significant violation of scale invariance, because the isocurvature perturbations are converted to curvature perturbations {\it before} inflation ends.  The curvature perturbations $\zeta$ are constant outside the horizon.  By converting $\sigma$ into $\zeta$ during inflation, the observable $\zeta$ modes are therefore insensitive to the end of inflation.  
\vskip 6pt
\hrule
\vskip 1pt
\hrule
\vskip 6pt
\normalsize 

\vskip 4pt
{\it Bispectrum.} \hskip 6pt The most intriguing aspect of the phenomenology of QSFI is the fact that it leads to large non-Gaussianities with a unique scaling behavior in the squeezed limit. As we alluded to in the Introduction, this provides the opportunities to use CMB and LSS measurements to probe Hubble-mass degrees of freedom during inflation. 

Chen and Wang~\cite{Chen:2009zp} explicitly computed the three-point correlation function induced by the interaction ${\cal L}_{\rm int} = - \mu \sigma^3$. In Appendix~\ref{sec:QSFI2}, we show how to reproduce the most important features of their answer from simple physical reasonings and back-of-the-envelope estimates.
For the amplitude of the bispectrum we find
\beq
{\cal L}_{\rm int} = - \mu \sigma^3  \qquad \Leftrightarrow \qquad  \includegraphicsbox[scale=.95]{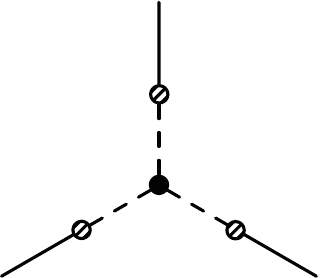} \qquad \Leftrightarrow \qquad f_{\rm NL} \sim \Delta_\zeta^{-1}\, \left( \frac{\rho}{H}\right)^3 \frac{\mu}{H}\ .
\eeq
Notice the enhancement of $f_{\rm NL}$ by the inverse of the amplitude of the primordial fluctuations, $\Delta_\zeta^{-1} \sim 10^5$. This allows for a large non-Gaussianity even in the perturbative regime with $\rho < H$ and $\mu < H$.
In Appendix~\ref{sec:QSFI2}, we also state simple ``Feynman rules" that allow us to estimate the size of more general diagrams. Essentially, each mixing insertion contributes a factor of $\rho/H$ and each interaction vertex gives a factor of $\mu/H$. The powers of $\Delta_\zeta^{-1}$ are determined by the number of $\sigma$'s in the interaction vertex, with interactions containing fewer $\sigma$'s being suppressed.

The $\sigma$ field is massive and decays outside of the horizon as $(-\tau)^{3/2-\nu}$, where $\nu \equiv \sqrt{\frac{9}{4} - \frac{m_\sigma^2}{H^2}}$ and $\tau$ is conformal time.
This leads to a non-trivial squeezed limit for the bispectrum
\beq
\lim_{k_1 \to 0} \langle \zeta_{{\bf k}_1} \zeta_{{\bf k}_2} \zeta_{{\bf k}_3}\rangle \ \propto\   \frac{1}{k_1^{{3/2} +\nu}}  \ . \label{equ:scaling}
\eeq
This scaling behavior is intermediate between that of the local shape ($k_1^{-3}$) and that of the equilateral shape ($k_1^{-1}$). By measuring the squeezed limit we determine the index $\nu$ and hence the mass of the isocurvaton $\sigma$.

The result (\ref{equ:scaling}) has a simple physical interpretation (see Appendix~\ref{sec:QSFI2}):
First, we recall that the squeezed limit corresponds to the correlation between a long-wavelength mode $k_1$ and two short-wavelength modes $k_2 \approx k_3$. 
The long mode crosses the horizon at $|k_1 \tau_1| \sim 1$, some time before the horizon crossing of the short modes at $|k_2 \tau_2| \sim 1$. The superhorizon evolution of the long mode between $\tau_1$ and $\tau_2$ leads to a suppression of its amplitude
\beq
\sigma_{k_1}(\tau_2) \ \sim\ \sigma_{k_1}(\tau_1) \left( \frac{\tau_2}{\tau_1}\right)^{3/2-\nu} 
\ \sim\  \sigma_{k_1}(\tau_1) \left( \frac{k_1}{k_2}\right)^{3/2-\nu} \ . \label{equ:time}
\eeq
We can write eq.~(\ref{equ:scaling}) in the following suggestive way,
\beq
\lim_{k_1 \to 0} \langle \zeta_{{\bf k}_1} \zeta_{{\bf k}_2} \zeta_{{\bf k}_3}\rangle \ \propto\   \underbrace{\frac{1}{k_1^3 k_2^3}}_{\rm local} \left( \frac{k_1}{k_2}\right)^{{3/2} -\nu} \ . 
\eeq
We recognize this  as the product of the local shape and a modulation that precisely matches the suppression of the long mode in eq.~(\ref{equ:time}).
This makes a lot of sense: if the isocurvaton is massless it freezes after horizon crossing and we expect local type non-Gaussianity from the non-derivative interaction $\sigma^3$. The squeezed limit of QSFI can therefore be interpreted as a {\it modulated local shape} with the modulation determined by the superhorizon evolution and hence the mass of the isocurvaton mode.

Future data \cite{Planck, SDSS} will have much to say about the primordial bispectrum in the squeezed limit. Here, we want to highlight the role of LSS observations.
In recent years, the
{\it scale-dependent bias} has emerged as a sensitive probe of primordial non-Gaussianity \cite{Dalal:2007cu}.
Receiving most of its signal from the squeezed limit of the bispectrum, the scale-dependent bias is an ideal probe of QSFI.
Eq.~(\ref{equ:scaling}) implies the following scaling for the non-Gaussian bias
\beq
\Delta b(k) \propto \frac{f_{\rm NL}}{k^{1/2+\nu}}\ .
\eeq
This prediction motivates generalizing the LSS data analysis to include $\nu$ as a free parameter; whereas, to date, the analysis has mostly been restricted to the case $\Delta b \propto k^{-2}$ \cite{Slosar:2008hx} (but see~\cite{Xia:2011hj, Shandera:2010ei}).

\vskip 4pt
{\it Trispectrum.} \hskip 6pt  QSFI makes further interesting predictions for the four-point function: 
First, we see that, correlated with the three-point function from the $\sigma^3$ interaction there is a four-point function from a scalar exchange diagram
\begin{align}
{\cal L}_{\rm int} = - \mu \sigma^3  &\qquad \Leftrightarrow \qquad \includegraphicsbox[scale=.75]{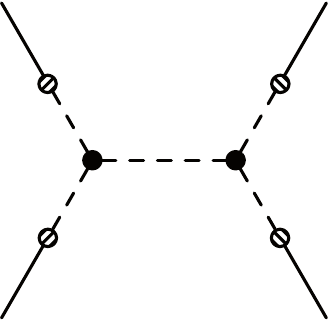} \qquad \Leftrightarrow \qquad \tau_{\rm NL} \sim \Delta_\zeta^{-2}\, \left( \frac{\rho}{H}\right)^4 \left(\frac{\mu}{H}\right)^2 \ \gg \ f_{\rm NL}^2\ . \label{equ:se} 
\end{align}
We note that QSFI gives a natural mechanism to boost the amplitude $\tau_{\rm NL}$ relative to $(\tfrac{6}{5} f_{\rm NL})^2$. (Recall that if a single source is responsible both for the power spectrum and the non-Gaussianity of $\zeta$ then $\tau_{\rm NL} = (\tfrac{6}{5} f_{\rm NL})^2$.) We expect this feature 
to lead to a {\it stochastic} halo bias (i.e.~bias inferred from $\langle \delta_{\rm h}  \delta_{\rm h} \rangle$ $\ne$ bias inferred from $\langle \delta_{\rm h}  \delta_{\rm m} \rangle$) \cite{Tseliakhovich:2010kf, Smith:2010gx}.
These signatures of QSFI deserve further investigation~\cite{Paper3}.

For models with additional quartic couplings such as $\sigma^4$, we get a four-point function associated with a contact interaction
\begin{align}
{\cal L}_{\rm int} = - \lambda \sigma^4  &\qquad \Leftrightarrow \qquad  \hspace{0.3cm} \includegraphicsbox[scale=.75]{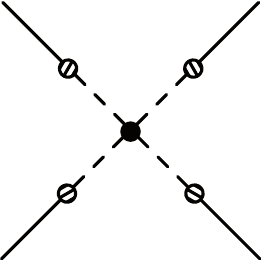}  \hspace{0.3cm} \qquad \Leftrightarrow \qquad g_{\rm NL} \sim \Delta_\zeta^{-2}\, \left( \frac{\rho}{H}\right)^4 \lambda \ . \label{equ:cont} 
\end{align}
The shapes of the four-point functions in (\ref{equ:se}) and (\ref{equ:cont}) are distinct.
The scalar exchange non-Gaussianity peaks in the collapsed configuration where all momenta have equal amplitude, while the contact interaction peaks in the squeezed configuration where one of the momenta vanishes.

\vskip 6pt
We will return to the phenomenological predictions of QSFI in future work~\cite{Paper3}. In this paper, we will instead explore the theoretical foundations of QSFI and its relation to supersymmetry.

\subsubsection*{Variants of Quasi-Single-Field Inflation}

The most general form of quasi-single-field inflation can be written as a deformation of slow-roll inflation:
${\cal L}_{\rm QSFI} =  {\cal L}_{\rm s.r.} + \Delta {\cal L}$, where 
\beq
\Delta {\cal L} = - \tfrac{1}{2} (\partial_\mu \sigma)^2 - \tfrac{1}{2} m_\sigma^2 \sigma^2 + {\cal L}_{\rm mix} + {\cal L}_{\rm int} \ .
\eeq
It is straightforward to extend the ideas of \cite{Chen:2009zp} to different mixing terms ${\cal L}_{\rm mix}$ and/or different interactions ${\cal L}_{\rm int}$.  
In this section, we will be systematic about this and discuss the range of possibilities in the effective theory approach. We begin with a non-SUSY treatment and then discuss its SUSY implementation.

\vskip 4pt
{\it Mixing terms.} \hskip 6pt Instead of ${\cal L}_{\rm mix}^{(1)} = m^3 [(\partial_\mu \varphi)^2 + 1] \sigma \to \rho \dot \pi_c \sigma$ we could consider the following mixing terms 
\begin{align}
{\cal L}_{\rm mix}^{(2)} &\ =\ \hat m^2_\alpha  [(\partial_\mu \varphi)^2 + 1] \partial_\mu \varphi \partial^\mu \sigma \ \, \to\ \alpha \hskip 1pt  \dot \pi_c \dot \sigma \ , \\
{\cal L}_{\rm mix}^{(3)} &\ =\ \hat m^2_\beta [\partial_\mu \varphi \partial^\mu \sigma - 3H \sigma] \ \ \ \ \ \ \to\ \beta  \hskip 1pt \partial_\mu \pi_c \partial^\mu \sigma \ ,
\end{align}
where $\{ \alpha, \beta\} \equiv \hat m^2_{\alpha, \beta}/(\Mp^2 |\dot H|)^{1/2}$.
However, it is easy to see that ${\cal L}_{\rm mix}^{(3)}$ does not lead to observable signatures.  Integrating by parts, we find ${\cal L}_{\rm mix}^{(3)} \to \hat m^2_\beta [-(\nabla^2 \pi) \sigma + \nabla^\mu (\nabla_\mu \pi \sigma) - 3 H \sigma ] $.  In the decoupling limit, $\dot H \to 0$, we have $\nabla^2 \pi = 0$ and therefore, the mixing term will not contribute.  Equivalently, we may remove the mixing term by a field redefinition $\pi_c \to \pi_c - \tfrac{\beta}{2} \sigma$.  Because $\sigma \to 0$ as $\tau \to 0$, such a field redefinition does not change the late-time correlation functions.  For these reasons, we may set $\beta = 0$ without loss of generality.\footnote{We thank Kendrick Smith for helpful discussions on these issues.}

Introducing ${\cal L}_{\rm mix}^{(2)}$ will modify the dispersion relations of physical modes.  We will require that no group velocity exceeds the speed of light.  Consider the Lagrangian
\beq
{\cal L}_{\rm kin} = \tfrac{1}{2} (\dot \pi_c \ \dot \sigma)  \left( \begin{array}{cc} 1 + \epsilon_\pi & \alpha \\ \alpha & 1 + \epsilon_\sigma \end{array} \right) \left( \begin{array}{c} \dot \pi_c  \\ \dot \sigma \end{array} \right) - \tfrac{1}{2} (\partial_i \pi_c \ \partial_i \sigma)  \left( \begin{array}{cc}  1\ & \ 0 \\ 0\ & \ 1 \end{array} \right) \left( \begin{array}{c} \partial_i \pi_c  \\ \partial_i \sigma \end{array} \right) \ ,
\eeq
where we have allowed for sound speeds for both $\pi$ and $\sigma$: i.e.~$c_{s,\pi}^{-2} \equiv 1 + \epsilon_\pi$ and $c_{s,\sigma}^{-2} \equiv 1 + \epsilon_\sigma$.  Solving the equations of motion, we find two positive frequency solutions with velocities given by
\beq
v_{\pm}= \frac{1}{(1+\epsilon_\sigma)(1+\epsilon_\pi) - \alpha^2}\left[1 + \tfrac{1}{2} (\epsilon_\sigma+ \epsilon_\pi) \pm \sqrt{ \tfrac{1}{4} (\epsilon_\sigma- \epsilon_\pi)^2+\alpha^2}\, \right] \ .
\eeq
The constraint $v_{+} \leq 1$ requires that $\epsilon_\sigma \epsilon_\pi \geq \alpha^2$.  If we take $\epsilon_\sigma = \epsilon_\pi = \epsilon$, the velocities simplify to $v_{\pm}  = (1 + \epsilon \pm \alpha)^{-1}$, which makes $\epsilon \geq \alpha$ transparent. In the perturbative regime, $\alpha < 1$, we only require a small deviation from the speed of light, $c \equiv 1$.
The fact that we have to introduce non-trivial sound speeds for both $\pi$ and $\sigma$ to avoid superluminal propagation is completely consistent with expectations from quantum field theory in flat space.  Note that sound speeds for $\pi$ and $\sigma$ arise from operators which include terms of the form $(\partial_{\mu} \pi \partial^\mu \pi)^2$, $(\partial_{\mu} \sigma \partial^\mu \pi)^2$, etc.  It was shown in \cite{Adams:2006sv}, that these types of four-derivative operators have coefficients that are constrained by unitarity to have non-zero values.  Although in the effective theory, it appears that we have to adjust these coefficients by hand, any manifestly Lorentz-invariant UV-completion of the effective theory will necessarily yield the coefficients consistent with the constraints.

\vskip 4pt
In summary, we will consider two mixing terms ${\cal L}_{\rm mix}^{(1)} = \rho \dot \pi_c \sigma$ and ${\cal L}_{\rm mix}^{(2)} = \alpha  \dot \pi_c \dot \sigma $.
For the mixing to be a perturbative effect, we require $\{ \frac{\rho}{H} , \alpha \} < 1$. Note that this is not a model-building requirement, and, in principle, we can allow the mixing parameters to be order one or larger~\cite{Paper3}.
 
\vskip 4pt
{\it Interactions.} \hskip 6pt 
These two quadratic mixing terms may be combined with any of the following cubic interactions
\begin{align}
\hat{\mathcal{L}}_{1a} &\ =\ m_a^3 [(\partial_\mu \varphi)^2+1]\, \sigma &\ \subset\ \ &  {\cal L}_{1a} =  m_a^3 (\partial_\mu \pi)^2  \sigma \label{equ:first}\\
\hat{\mathcal{L}}_{1b} &\ =\ m_b^3 (\partial_\mu \varphi)^2 [ (\partial_\mu \varphi)^2+1]\, \sigma &\ \subset\ \ &  {\cal L}_{1b} =  m_b^3 \left(\dot \pi^2 \sigma - (\partial_\mu \pi)^2  \sigma \right) \\
\hat{\mathcal{L}}_2 &\ =\  \hat m^2 [(\partial_\mu \varphi)^2+1]\, \partial_\mu \varphi \partial^\mu \sigma &\ \subset\ \ & \ {\cal L}_2 =  \hat m^2\left( (\partial_\mu \pi)^2 \dot \sigma + \dot \pi \partial_\mu \pi \partial^\mu \sigma \right) \\
\hat{\mathcal{L}}_3 &\ =\  \tilde m^2 (\partial_\mu \varphi)^2 \sigma^2 &\ \subset\ \ & \ {\cal L}_3 =  \tilde m^2  \dot \pi \sigma^2 \\
\hat{\mathcal{L}}_{4a} &\ =\  \bar m_a  (\partial_\mu \varphi \partial^\mu \sigma) \sigma  &\ \subset\ \ & {\cal L}_{4a}= \bar m_a \partial_\mu \pi \partial^\mu \sigma \sigma \\
\hat{\mathcal{L}}_{4b} &\ =\ \bar m_b (\partial_\mu \varphi)^2  (\partial_\mu \varphi \partial^\mu \sigma) \sigma  &\ \subset\ \ & {\cal L}_{4b} = \bar m_b \left( \partial_\mu \pi \partial^\mu \sigma \sigma + 2 \dot \pi \dot \sigma \sigma \right) \\
\hat{\mathcal{L}}_{5a} &\ =\  \lambda_a  (\partial_\mu \varphi)^2 (\partial_\mu \sigma)^2 &\ \subset\ \ & {\cal L}_{5a} = \lambda_a  \dot \pi (\partial_\mu \sigma)^2 \\
\hat{\mathcal{L}}_{5b} &\ =\  \lambda_b  (\partial_\mu \varphi \partial^\mu \sigma)^2 &\ \subset\ \ & {\cal L}_{5b}= \lambda_b  \partial_\mu \pi  \partial^\mu \sigma \dot \sigma \\
\hat{\mathcal{L}}_{5c} &\ =\  \lambda_c  (\partial_\mu \varphi)^2 (\partial_\mu \varphi \partial^\mu \sigma)^2 &\ \subset\ \ &{\cal L}_{5c} = \lambda_c \left( \dot \pi \dot \sigma^2 - \partial_\mu \pi \partial^\mu \sigma \dot \sigma \right) \\
\hat{\mathcal{L}}_6 &\ =\ \mu \sigma^3 &\ \subset \ \ & \ {\cal L}_6 = \mu \sigma^3\\
\hat{\mathcal{L}}_7 &\ =\ \lambda (\partial_\mu \varphi \partial^\mu \sigma)  \sigma^2 &\ \subset\ \ & \ {\cal L}_7= \lambda \dot \sigma \sigma^2 \\
\hat{\mathcal{L}}_{8a} &\ =\ \Lambda_1^{-1} \sigma (\partial_\mu \sigma)^2 &\ \subset \ \ & {\cal L}_{8a}= \Lambda_1^{-1} \sigma (\partial_\mu \sigma)^2  \\
\hat{\mathcal{L}}_{8b} &\ =\ \Lambda_2^{-1}  \sigma (\partial_\mu \varphi \partial^\mu \sigma)^2  &\ \subset\ \ &{\cal L}_{8b} = \Lambda_2^{-1}  \sigma  \dot \sigma^2 \\
\hat{\mathcal{L}}_{9a} &\ =\ \Lambda_3^{-2} (\partial_\mu \varphi \partial^\mu \sigma) (\partial_\mu \sigma)^2  &\ \subset\ \ & {\cal L}_{9a} =  \Lambda_3^{-2} \dot \sigma (\partial_\mu \sigma)^2 \\
\hat{\mathcal{L}}_{9b} &\ =\ \Lambda_4^{-2} (\partial_\mu \varphi \partial^\mu \sigma)^3  &\ \subset\ \ & {\cal L}_{9b} = \Lambda_4^{-2} \dot \sigma^3  \label{equ:last}
\end{align}
This list captures all possible combinations of the fields $\varphi$ and $\sigma$.  We have restricted to operators with at most one derivative acting on each field, but one may easily include additional derivatives.
In Appendix~\ref{sec:QSFI2}, we show how to estimate the size of non-Gaussianities for these interactions. The results are summarized in Table~\ref{tab:NG}. In the table we also indicate which interactions, in principle, allow for large non-Gaussianities ({\bf large NG}), which peak in the squeezed limit ({\bf S.L.}), which preserve supersymmetry ({\bf SUSY}), and which ultimately lead to natural models ({\bf Natural}).

\vskip 4pt
In the remainder of this section we will explore which of these new models of QSFI have a natural microphysical implementation.
We will distinguish cases that are strictly unnatural because they involve fine-tuning at {\it tree level} (\S\ref{sec:tree}); cases whose naturalness remains to be established because they are likely to  involve fine-tuning at {\it loop level} (\S\ref{sec:loop}); and finally cases that are completely natural because supersymmetry regulates the radiative corrections (\S\ref{sec:SUSY_QSFI}).

 \begin{table}[h!]
\caption{\sl Non-Gaussianity in Quasi-Single Field Inflation. }
\label{tab:NG}
\vspace{-0.5cm}
\begin{center}
\begin{tabular}{l  l  l   c c c c}
\toprule
\hspace{0.2cm} {\footnotesize \bf Interaction} & $f_{\rm NL}^{(1)}$ & $f_{\rm NL}^{(2)}$  & {\footnotesize \bf Large NG}  & {\footnotesize \bf S.L.} &  {\footnotesize \bf SUSY} & {\footnotesize \bf Natural}\\
\otoprule
$\ {\cal L}_{1a} = m_a^3 (\partial_\mu \pi)^2  \sigma$ &  $\left(\frac{\rho}{H} \right)^2 $ & $\alpha \, \frac{\rho}{H}$ & &  $\checkmark$ &  $\checkmark$ & \\
 \midrule
  $\  {\cal L}_{1b} = m_b^3 (\dot \pi)^2  \sigma$ & $\left(\frac{\rho}{H} \right)^2$ & $\alpha\, \frac{\rho}{H}$  & &  $\checkmark$ &  $\checkmark$ &\\
 \midrule 
 $\ {\cal L}_2 = \hat m^2  (\partial_\mu \pi)^2 \dot \sigma $ & $\frac{\rho}{H}\, \alpha$ & $\alpha^2$  & & &\\
 \midrule
 $\ {\cal L}_3 = \tilde m^2  \dot \pi \sigma^2 $ & $ \left(\frac{\rho}{H}\right)^2 \left(\frac{\tilde m}{H} \right)^2    $ & $ \alpha^2 \left(\frac{\tilde m}{H} \right)^2   $ &    &  $\checkmark$ & & \\
 \midrule
 ${\cal L}_{4a} = \bar m_a \partial_\mu \pi \partial^\mu \sigma \sigma$ & $\left(\frac{\rho}{H}\right)^2 \frac{\bar m_a}{H}$ & $\alpha^2 \,\frac{\bar m_a}{H}$ &   &  $\checkmark$ &  $\checkmark$ &\\
  \midrule
 ${\cal L}_{4b} = \bar m_b \dot \pi \dot \sigma \sigma$ & $\left(\frac{\rho}{H}\right)^2 \frac{\bar m_b}{H}$ & $\alpha^2 \,\frac{\bar m_b}{H}$ &   &  $\checkmark$ & &\\
 \midrule
 ${\cal L}_{5a} = \lambda_a  \dot \pi (\partial_\mu \sigma)^2$ & $ \left( \frac{\rho}{H}\right)^2 \lambda_a$ & $ \alpha^2\, \lambda_a$ &   & &\\
 \midrule ${\cal L}_{5b} = \lambda_b  \partial_\mu \pi \partial^\mu \sigma \dot \sigma  $ & $ \left( \frac{\rho}{H}\right)^2 \lambda_b$ & $ \alpha^2\, \lambda_b$ &   & &\\
 \midrule 
 ${\cal L}_{5c} = \lambda_c \dot \pi \dot \sigma^2$ & $ \left( \frac{\rho}{H}\right)^2 \lambda_c$ & $ \alpha^2\, \lambda_c$ &  & &\\
 \midrule
 $\ {\cal L}_6 =  \mu \sigma^3$ & $\left(\frac{\rho}{H}\right)^3 \frac{\mu}{H} \, \Delta_\zeta^{-1}  $ & $\alpha^3\, \frac{\mu}{H} \, \Delta_\zeta^{-1}  $  & $\checkmark$  &  $\checkmark$ &  $\checkmark$ & {\Large $\checkmark$} \\
 \midrule
 $\ {\cal L}_7 = \lambda \dot \sigma \sigma^2 $ & $ \left( \frac{\rho}{H} \right)^3 \lambda \, \Delta_\zeta^{-1} $ &  $ \alpha^3\, \lambda \,  \Delta_\zeta^{-1} $ & $\checkmark$  &  $\checkmark$ & & (?)  \\
 \midrule
 ${\cal L}_{8a} = \Lambda_1^{-1} (\partial_\mu \sigma)^2 \sigma $ & $ \left( \frac{\rho}{H}\right)^3 \frac{H}{\Lambda_1}\,  \Delta_\zeta^{-1}$ & $ \alpha^3\, \frac{H}{\Lambda_1}\,  \Delta_\zeta^{-1}$ & $\checkmark$  &  $\checkmark$  &  $\checkmark$ & {\Large $\checkmark$} \\
 \midrule ${\cal L}_{8b} = \Lambda_2^{-1}  \dot \sigma^2  \sigma $ & $  \left( \frac{\rho}{H}\right)^3 \frac{H}{\Lambda_2} \, \Delta_\zeta^{-1}$ & $  \alpha^3\, \frac{H}{\Lambda_2} \, \Delta_\zeta^{-1}$ & $\checkmark$  &  $\checkmark$ & & (?) \\
 \midrule
 ${\cal L}_{9a} = \Lambda_3^{-2} \dot \sigma (\partial_\mu \sigma)^2$ & $  \left( \frac{\rho }{H}\right)^3  \big(\frac{H}{\Lambda_3} \big)^2\, \Delta_\zeta^{-1}$ & $  \alpha^3\,  \big(\frac{H}{\Lambda_3} \big)^2\, \Delta_\zeta^{-1}$ & $\checkmark$ & & & $\checkmark$ \\
 \midrule ${\cal L}_{9b} = \Lambda_4^{-2} \dot \sigma^3 $ & $ \left( \frac{\rho }{H}\right)^3 \big( \frac{H}{\Lambda_4}\big)^2\, \Delta_\zeta^{-1}$ & $ \alpha^3\, \big( \frac{H}{\Lambda_4}\big)^2\, \Delta_\zeta^{-1}$ & $\checkmark$ & & & $\checkmark$ \\
\bottomrule
\end{tabular}
\end{center}
\end{table}

\subsection{Problems at Tree Level}
\label{sec:tree}

When organizing the effective theory of inflation in \S\ref{sec:adi}, we found it convenient to expand the action in terms of operators like $[(\partial_\mu \varphi)^2 + 1]$.  
However, when discussing the natural values of parameters, one should think of $(\partial_\mu \varphi)^2$ and $1$ as independent operators.  In general, writing terms like $[(\partial_\mu \varphi)^2 + 1] {\cal O}(\varphi)$, without independently adding the operator ${\cal O}(\varphi)$ should be viewed as a tree-level fine-tuning.\footnote{The fine-tuning of terms necessary to cancel tadpoles is an exception to this rule about naturalness.  In writing the EFT of inflation, we demanded that our background solution takes a specific form and then fix certain coefficients to cancel tadpoles.  Tadpole cancelation is not a tuning of the action, but simply represents the freedom to choose initial conditions. However, beyond the cancelation of tadpoles, the cancelation of the constant in $(\partial_\mu \varphi)^2 = -1 + \cdots$ will typically require tree-level fine-tuning. In some exceptional cases, there may be a dynamical explanation for the cancelations.  However, in general, we should be suspicious of exact cancelations between operators. }  This is most apparent in the context of supersymmetry, where the operators proportional to $1$ and $(\partial_\mu \varphi)^2$ typically arise from fundamentally different terms in the K\"ahler or superpotential, cf.~\S\ref{sec:SUSYPX}.  Cancelling the constant contribution from $(\partial_\mu \varphi)^2$ only occurs when the coefficients of a priori independent operators are carefully chosen to cancel. With this in mind we wrote the operators in eqs.~(\ref{equ:first})--(\ref{equ:last}) without the combination $[(\partial_\mu \varphi)^2 + 1]$, unless it was related to tadpole cancellation in either $\pi$ or $\sigma$.

It then follows immediately that the interactions ${\cal L}_1$ -- ${\cal L}_5$ can't lead to large non-Gaussianities, unless the {\it tree-level action is fine-tuned}. Let us demonstrate this case-by-case:
\begin{itemize}
\item The interactions ${\cal L}_{1a,b}$ come from the same operator $\hat{\cal L}_1$ that leads to the mixing term $\rho \dot \pi_c \sigma$, where $\rho \equiv m_{a,b}^3/(\Mp^2 |\dot H|)^{1/2}$. We therefore can't make the interaction large without inducing a large mixing term. This constrains the non-Gaussianity to be small, $f_{\rm NL} < 1$.  For ${\cal L}_{1b}$, one may decouple the interaction and mixing terms, but this requires fine tuning.
\item Similarly, the interaction ${\cal L}_2$ comes from the same operator that leads to the mixing term $\alpha\, \dot \pi_c \dot \sigma$. Again, this constrains the non-Gaussianity to be small, $f_{\rm NL} < 1$.
\item The interaction ${\cal L}_3$ could give large non-Gaussianity for $\tilde m \gg H$. However, the operator $\hat{\cal L}_3  = \tilde m^2 (\partial_\mu \varphi)^2 \sigma^2 \subset \tilde m^2 \sigma^2$ also contains a tree-level mass term for $\sigma$. Without fine-tuning, we require $\tilde m^2 \lesssim H^2$ and hence $f_{\rm NL} < 1$.
\item Similarly, the interactions ${\cal L}_{4a,b}$ are necessarily related to the operator $\bar m_{a,b}\dot \sigma \sigma \to H \bar m_{a,b} \sigma^2$. Keeping the mass naturally small requires $m_{a,b} < H$ and forces the non-Gaussianity to be small, $f_{\rm NL} < 1$.
\item The interactions ${\cal L}_{5a,b,c}$ only lead to large non-Gaussianities if $\lambda_{a,b,c} \gg 1$. Again, it is straightforward to see that this requires fine-tuning. The interactions ${\cal L}_{5a,b,c}$ come from operators that also contain a contribution to the kinetic term of the form $\lambda_{a,b,c} \dot \sigma^2$.  Without fine tuning, if $\lambda_{a,b,c} \gg 1$ the true canonical field is $\sigma_c \sim \sigma / \lambda^{1/2}$ and  $f_{\rm NL} \ll 1$.  
\end{itemize}
On these grounds we reject the interactions ${\cal L}_1$ -- ${\cal L}_5$ as candidates for natural models of QSFI.
For the remaining interactions naturalness is a bit less trivial to check.

\subsection{Challenges at Loop Level}
\label{sec:loop}

{\it Naturalness criteria.} \hskip 6pt 
Given an EFT with a set of couplings, one should ask if these parameters are stable under radiative corrections. Loop contributions to various parameters can be estimated as a function of the UV-cutoff of the theory.  If a parameter receives large corrections, we call the parameter fine tuned.
In many scenarios, the UV-cutoff of the EFT is unknown and, in principle, may be taken to much higher energies than are probed directly in experiments.  Parameters may become unnatural if the UV-cutoff (and hence the range of validity of the EFT) is taken to be too large.  One may also reverse the logic to predict the breakdown of the EFT based on naturalness.




Before discussing supersymmetry, it will be useful to first understand the naturalness of QSFI with minimal field content and interactions. 
 Models based on the interactions ${\cal L}_6$ and ${\cal L}_7$ are effectively described by a slow-roll background with weakly-coupled interactions, so, in principle, no new physics is required up to the symmetry breaking scale $\Lambda_b \simeq (\Mp^2 |\dot H|)^{1/4}$~\cite{Baumann:2011su}.  In that case, the UV-cutoff of the effective theory can be as large as $\Lambda_{\rm uv} \sim \Lambda_b$.
 On the other hand, models based on ${\cal L}_8$ and ${\cal L}_9$ may become strongly coupled at energies below the symmetry breaking scale. As we discussed in \S\ref{sec:weak}, this can be viewed as an indication that new physics should appear at or below that strong coupling scale, $\Lambda_\star = \Lambda_i$. Hence, for models based on ${\cal L}_8$ and ${\cal L}_9$ we impose $\Lambda_{\rm uv} = {\rm min}\{ \Lambda_b , \Lambda_i \}$.
 
 The absence of any additional new physics up to $\Lambda_b$ or $\Lambda_\star$ leads to a relatively strong condition for naturalness: namely, the theory ought to be natural for $\Lambda_{\rm uv} \sim {\rm min}\{ \Lambda_b, \Lambda_\star\} \gg H$.  However, of course, we can't exclude the possibility that new physics may appear below the scale ${\rm min}\{ \Lambda_b, \Lambda_\star\} $ and that this would improve the radiative stability of the models.  Ultimately, SUSY will play this role for some of the interactions.  

\vskip 4pt
{\it Loop corrections.} \hskip 6pt 
A principle worry is that the large interactions required for observable non-Gaussianities induce large loop corrections to the mass of the isocurvaton:
\beq
\delta m_\sigma^2 \ =\  \includegraphicsbox[scale=0.9]{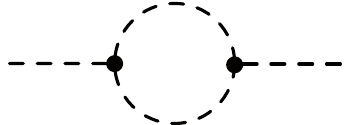}\ . \eeq
For the QSFI mechanism to be operative, we require $m_{\sigma}^2 \leq \tfrac{9}{4} H^2$.  A necessary condition for naturalness is therefore $\delta m_\sigma^2 \lesssim  \tfrac{9}{4} H^2$. Let us study the interactions ${\cal L}_6$ -- ${\cal L}_9$ one-by-one:

\begin{itemize}
\item The interaction ${\cal L}_6$, is somewhat special, in the sense that loop contributions to the mass are {\it finite}
\beq
\delta m_{\sigma \{6\}}^2 \sim \mu^2\ .
\eeq
This is smaller than the supergravity-induced mass $m_\sigma^2 \sim H^2$, if $\mu \lesssim H$. Hence, ${\cal L}_6$ is a promising candidate for supersymmetric QSFI. Indeed, in \S\ref{sec:SUSY_QSFI}, we will present a natural model based on ${\cal L}_6$.

\item Without SUSY, the interactions ${\cal L}_7$ -- ${\cal L}_9$ all give {\it UV-divergent} loop corrections:

- The interaction ${\cal L}_7$  induces the following one-loop mass renormalization
\begin{align}
\delta m_{\sigma\{7\}}^2 &\sim \lambda^2 \Lambda_{\rm uv}^2 \ .
\end{align}
In the absence of additional new physics below the symmetry breaking scale, we cut off the loop at $\Lambda_{\rm uv} \sim \Lambda_b \sim (\Mp^2 |\dot H|)^{1/4}$.
Requiring the loop contribution to be subdominant relative to the tree-level contribution, $m_\sigma \sim H$, puts a bound on the size of the non-Gaussianity
\begin{align}
f_{\rm NL \{7\}} &< \alpha^3 \Delta_\zeta^{-1/2} \ .
\end{align}
This allows marginally observable non-Gaussianity, but only under the optimistic assumption $\alpha \sim 1$.

- The interactions ${\cal L}_8$ leads to the following radiative contribution to the mass
\begin{align}
\delta m_{\sigma\{8\}}^2 &\sim  \frac{\Lambda_{\rm uv}^4}{\Lambda^2_{1,2}} \ , \label{equ:loop8}
\end{align}
where $\Lambda_{1,2} > H$ in order for the theory to be weakly coupled at horizon exit.
The naturalness condition $\delta m_{\sigma\{8\}} < H$, therefore requires $\Lambda_{\rm uv} < \Lambda_{1,2}$.  This cutoff is only consistent with $\Lambda_{\rm uv} = {\rm min}\{ \Lambda_b , \Lambda_{1,2} \}$ for $\Lambda_{1,2} > \Lambda_b$ and $\Lambda_{\rm uv} \sim \Lambda_b$.
Naturalness then strongly constrains the non-Gaussianity associated with ${\cal L}_8$,
\begin{align}
f_{\rm NL \{8\}} &< \alpha^3 \ .
\end{align}
Hence, without a way to regularize the loops below the symmetry breaking scale, the interactions ${\cal L}_{8a,b}$ don't naturally lead to observable non-Gaussianities. 
Alternatively, we can keep a more open mind about the possibility of new physics appearing at relatively low energies. To achieve a given $f_{\rm NL}$ with natural parameters, one then finds a bound on the UV-cutoff
\begin{align}
\Lambda_{{\rm uv}\{8\}} &< \frac{\alpha^{3/2}}{f_{\rm NL}^{1/2}} \cdot (\Mp^2 |\dot H|)^{1/4} \ .
\end{align}

- The interactions ${\cal L}_9$ leads to wavefunction renormalization
\begin{align}
\delta Z_{\sigma \{9\}} &\sim \frac{\Lambda_{\rm uv}^4}{\Lambda_{3,4}^4} \ . \label{equ:lastX}
\end{align}
For $\Lambda_{\rm uv} \sim {\rm min}\{\Lambda_b, \Lambda_{3,4}\}$, this is always smaller than the tree-level value $Z_\sigma = 1$.
The interactions ${\cal L}_9$ can therefore lead to natural models with large $f_{\rm NL}$. However, being pure derivative interactions, the non-Gaussianity is suppressed in the squeezed limit, making these models less interesting for our present purposes.

\end{itemize}

\subsection{Supersymmetry and Naturalness}
\label{sec:SUSY_QSFI}

Supersymmetry plays two important roles in the microphysical implementations of quasi-single-field inflation:
\begin{enumerate}
\item SUSY naturally and inevitably introduces at least one additional scalar field with mass of order the Hubble scale $H$.
\item SUSY can help to regulate dangerous loop corrections that would otherwise become important in the limit of large interactions and threaten the naturalness of the model.
\end{enumerate}
Can SUSY help to regulate the UV-divergences that we found in the previous section? Only for~${\cal L}_{8a}$. It is easy to understand why:
as we discussed in \S\ref{sec:break}, spontaneous breaking of Lorentz invariance necessarily breaks supersymmetry.  Therefore, any interaction that is not manifestly Lorentz invariant must be proportional to SUSY breaking.  This is the case for the cubic interactions $\mathcal{L}_{7}$, ${\cal L}_{8b}$ and ${\cal L}_{9a,b}$.  Therefore, when we embed these interactions in a supersymmetric theory, there will be no extra cancelations below the SUSY breaking scale, namely $\omega < \Lambda_b = (\Mp^2 |\dot H|)^{1/4}$.  
What distinguishes $\mathcal{L}_{8a}$ is that this interaction does not know about SUSY breaking directly.  The same interaction can occur in theories with unbroken SUSY.  Because the mass for $\sigma$ arises from SUSY breaking, it will be further suppressed relative to the estimate in (\ref{equ:loop8}).  
Although the interaction ${\cal L}_{8a}$ preserves SUSY, the mixing terms $\rho \dot \pi_c \sigma$ and $\alpha \dot \pi_c \dot \sigma$ don't.
A contribution to the mass of $\sigma$ is therefore generated by inserting the mixing terms inside the loop.\footnote{This is the same effect that we discussed in \S\ref{sec:weak}; see~eq.~(\ref{equ:loop1}).} The associated loops are still divergent, but suppressed relative to (\ref{equ:loop8}).
For the mixing $\alpha \dot \pi_c \dot \sigma$, the suppression is rather mild, since the order of the divergence stays the same, 
\beq
\delta m_{\sigma \{8a, \alpha\}}^2 \ =\ \sigma \includegraphicsbox[scale=1.0]{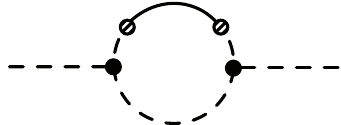}\ \sigma \ \sim\  \frac{\alpha^2}{\Lambda_1^2} \, \Lambda_{\rm uv}^4\ . 
\eeq
In the regime of interest, $\Lambda_1 < \Lambda_b$ and hence $\Lambda_{\rm uv} \sim \Lambda_1$. Naturalness, $\delta m_{\sigma \{8a, \alpha\}} \sim \alpha \Lambda_1 < H$, then puts a bound on the size of non-Gaussianity
\beq
f_{\rm NL\{8a,\alpha\}} < \left( \frac{H}{\Lambda_1}\right)^4 \Delta_\zeta^{-1}\ .
\eeq
This could marginally be observable, but only in the questionable limit that $\Lambda_1$ get dangerously close to $H$.

It is more promising to work with the mixing term $\rho \dot \pi_c \sigma$.
In this case, the order of the divergence is reduced,
\beq
\delta m_{\sigma \{8a, \rho \}}^2 \ =\ \sigma \includegraphicsbox[scale=1.0]{L8}\ \sigma \ \sim\  \frac{\rho^2}{\Lambda_1^2} \, \Lambda_{\rm uv}^2\ . 
\eeq
This suppression is sufficient to make the model natural.
In particular, for $\Lambda_{\rm uv} \sim {\rm min}\{\Lambda_b, \Lambda_1\}$, we find $\delta m_{\sigma \{8a\}} \lesssim \rho < H$.
Naturalness therefore doesn't put an additional constraint on the size of non-Gaussianity.
In particular we can get large non-Gaussianity even for $\Lambda_1$ safely above $H$,
\beq
f_{\rm NL \{8a,\rho\}} < \frac{H}{\Lambda_1} \Delta_{\zeta}^{-1}\ .
\eeq
Even for $\Lambda_1 \to \Lambda_b$, the non-Gaussianity can be observably large, $f_{\rm NL \{8a,\rho\}} < \Delta_{\zeta}^{-1/2}$.
The interaction ${\cal L}_{8a}$ is therefore a promising candidate for a natural SUSY model of QSFI.

\vskip 4pt
Although SUSY does not regulate the divergences associated with $\mathcal{L}_{7}$ and ${\cal L}_{8b}$, there is no direct constraint on $f_{\rm NL}$ if the cutoff is lowered.  We may have still have natural models for these interactions, but it would require new physics to enter at low energies.  We distinguish these situations from `natural SUSY implementations of QSFI', as they require something beyond SUSY (or in place of SUSY) to be natural.  
\vskip 8pt
We have identified two interactions ${\cal L}_{\rm int} = \{{\cal L}_6, {\cal L}_{8a}\}$ with the potential to lead to  natural models under minimal assumptions. Next, we will discuss whether these models can indeed be consistently embedded in a supersymmetric framework.

\subsubsection*{Model I}

We first present a SUSY implementation of QSFI using the interaction ${\cal L}_6$.  In the interest of economy, we would like to find a model where the massive field, $\sigma$, is part of the same chiral multiplet as $\varphi$.  This choice restricts the form that the model can take.  We can't obtain ${\cal L}_6$ from a superpotential, since that would break the shift symmetry $\varphi \to \varphi+const$.  Derivative mixings, such as $\partial_\mu \pi_c \partial^\mu \sigma$, are also difficult to generate, making it most natural to consider the mixing term $\rho \dot \pi_c \sigma$.

\vskip 4pt
Making these choices, the SUSY implementation of the model with ${\cal L}_{\rm mix} = \rho \dot \pi_c \sigma$ and ${\cal L}_{\rm int} = {\cal L}_6=-\mu \sigma^3$ is
\beq
\label{equ:model1}
{\cal L}_{\rm QSFI}^{\{{\rm I}\}} = \int d^4 \theta \, \Big[\, \tfrac{1}{2}(\G+\Gd)^2 \, +\, \underbrace{\frac{1}{\Lambda_1} (\G + \Gd)^3 }_{K_{\rm mix}} \, + \, \underbrace{\frac{1}{(\Lambda_2)^3} (\G + \Gd)^5}_{K_{\rm int}}\, \Big] \, +\, {\cal L}_X\ ,
\eeq
with
\beq
\rho \equiv \frac{(\Mp^2 |\dot H|)^{1/2}}{\Lambda_1} \qquad {\rm and} \qquad
\mu \equiv \frac{\Mp^2 |\dot H|}{(\Lambda_2)^3}\ .
\eeq
The constraints $\rho \lesssim H$ and $\mu \lesssim H$, require $\Lambda_1 \gtrsim \sqrt{\varepsilon} M_{\rm pl}$ and $\Lambda_2 \gtrsim \varepsilon^{1/3}(\frac{H}{M_{\rm pl}})^{1/3} M_{\rm pl} \sim  (\Delta_\zeta)^{1/3} \sqrt{\varepsilon} M_{\rm pl}$.
We see that Planck-suppressed corrections are sufficient to generate the required couplings.
An alternative way to generate the cubic coupling is the operator $K_{\rm int} = \frac{1}{(\Lambda_2)^3} (\G + \Gd)^3 X^\dag X$. In this case $\mu \sim \Mp^2 H^2/(\Lambda_2)^3$, which implies $\mu \lesssim H$ if $\Lambda_2 \gtrsim (\varepsilon \Delta_\zeta^2)^{1/6} \Mp$. 
We have not written an explicit mass term for $\sigma$ since it will be generated by supergravity corrections: $m_\sigma \sim H$.  Supergravity will also generate a cubic term, but with a coefficient that is suppressed by $\Lambda_1$, namely $\mu \sim 3H^2/\Lambda_1 \ll H$.

As we noted above, even in the absence of SUSY the mass of $\sigma$ is does not receive dangerously large corrections, if $\rho < H$ and $\mu < H$.  Embedding the model in supersymmetry does not change this situation.  Explicit loop calculations in the supersymmetric model simply reproduce the results of the previous section.  The role of SUSY in this model is to provide an explanation for the extra Hubble-mass scalar $\sigma$.  

\subsection*{Model II} 

Next, we discuss a natural SUSY implementation of our second candidate model, ${\cal L}_{8a}$.  As in the previous model, it would be ideal to embed $\sigma$ in a single chiral multiplet with $\varphi$.  We will start with this assumption, only to find it cannot lead to large $f_{\rm NL}$.  However, we will show that adding a second chiral field, $\Sigma$, leads to natural models with large non-Gaussianity.

\vskip 4pt
In order to generate $f_{\rm NL} > 1$ from the interaction ${\cal L}_{8a}$, we require 
\beq
\label{equ:CONX}
\Lambda_1 \ll H \Delta_\zeta^{-1} \sim \Delta_\zeta^{-1/2}(\Mp^2 |\dot H|)^{1/4}\ .
\eeq
In this case, SUSY is required to regulate the loop correction to the mass of $\sigma$.
Indeed, we can write the operator ${\cal L}_{8a}$ supersymmetrically
\beq
\label{equ:L2}
\int d^4 \theta\, \frac{1}{\Lambda_1} (\G+\Gd)^3 \subset \frac{1}{\Lambda_1} \sigma (\partial_\mu \sigma)^2\ .
\eeq
Since SUSY is unbroken by the interaction, the previously dangerous loop can now be controlled.
However, eq.~(\ref{equ:L2}) necessarily includes the mixing term
\beq
{\cal L}_{\rm mix} = \frac{\Mp^2 \dot H}{\Lambda_1} \dot \pi \sigma\ .
\eeq
With the constraint (\ref{equ:CONX}), this leads to $\rho \dot \pi_c \sigma$, with $\rho \gg H$. This spoils the QSFI mechanism, since large $\rho$ implies a large effective mass for $\sigma$.
We therefore find that QSFI with interaction ${\cal L}_{8a}$ can't be realized in a minimal setup with only a single chiral superfield $\Phi$.

It is clear that this obstruction to naturalness is specific to the case of a single chiral field.
Next, we explore the possibility of realizing the model with additional chiral fields.  It is now straightforward to separate the mixing term $\rho \dot \pi_c \tilde \sigma$ from the interaction $\mathcal{L}_{8a}$.
It is easy to see that the following Lagrangian leads to the desired couplings,
\beq
\label{equ:model2}
{\cal L}_{\rm QSFI}^{\{{\rm II}\}} = \int d^4 \theta \, \Big[ \,  \tfrac{1}{2}(\G+\Gd)^2 \, +\, \Sigma \Sigma^\dag \,+\, \underbrace{\frac{1}{\tilde \Lambda} (\G + \Gd)^2(\Sigma + \Sigma^\dag) }_{K_{\rm mix}} \, + \, \underbrace{\frac{1}{\Lambda_1} (\Sigma + \Sigma^\dag) \Sigma \Sigma^\dag}_{K_{\rm int}}\, \Big] \, +\, {\cal L}_X \ .
\eeq
This action contains a tadpole for $\tilde \sigma$ that we are implicitly canceling through an additional small coupling to $X$.  The spectator field $\Sigma$ does not break SUSY by itself, but only indirectly through its coupling to the inflaton $\Phi$.  In the limit $\tilde \Lambda \to \infty$, the action for $\Sigma$ decouples and describes a supersymmetric subsector (prior to coupling to supergravity, as usual).  By the usual SUSY non-renormalization theorems, we will not generate a mass for $\sigma$ through any loop diagram.
For example, if we were to insert $(K_{\rm int})^2$ and compute the loop, we would indeed find a divergence.  However, the divergence is only logarithmic and contributes to the wavefunction renormalization  $ Z \int d^4 \theta\, \Sigma \Sigma^{\dagger}$ rather than to the mass term.  By inserting $(K_{\rm mix})^2$, we can introduce a mass for $\tilde \sigma$, but this mass is smaller than Hubble when we cut off the loop at $\Lambda_{\rm uv} \sim {\rm min}\{ \Lambda_b, \Lambda_1\}$.
The theory described by eq.~(\ref{equ:model2}) is therefore technically natural.

\subsection{Summary}

We summarize our conclusions regarding the naturalness of quasi-single-field inflation in the presence of supersymmetry:

\begin{itemize}
\item ${\cal L}_1$ -- ${\cal L}_5$ are unnatural and require fine-tuning at tree-level.

\item ${\cal L}_7$ and  ${\cal L}_{8b}$ require additional physics beyond SUSY for naturalness.

\item ${\cal L}_{9a,b}$ are natural, but the signal is suppressed in the squeezed limit.

\item ${\cal L}_6$ is natural. Loops are finite even in the absence of SUSY, but SUSY explains the origin of the massive isocurvaton. The model requires only a single chiral multiplet that contains both the inflaton and the isocurvaton.

\item ${\cal L}_{8a}$ is natural, but only if the inflaton and the isocurvaton live in separate multiplets. SUSY is required to explain radiative stability.

\end{itemize}

\section{Conclusions}
\label{sec:conclusions}

All of the current CMB and LSS data is completely consistent with the simplest effective field theory of inflation---a theory of Goldstone bosons of spontaneously broken time translations~\cite{Cheung}.
However, despite this phenomenological success, the fundamental origin of the inflationary expansion remains a mystery.
The basic challenge is easy to understand: it is hard to protect fundamental scalar fields from radiative corrections.
The simplest field theory models of inflation are therefore unnatural in the technical sense~\cite{'tHooft:1979bh}.
Taking this naturalness problem seriously, one is led to consider symmetries as a way of protecting the inflationary background from destructive quantum corrections. This means either internal global symmetries or supersymmetry. 
The role of global symmetries in inflationary models is well-understood, although generic Planck-scale breaking effects are rarely included~\cite{Baumann:2010ys,Baumann:2010nu}.
Unfortunately, it seems that models with global symmetries are  observationally indistinguishable from models that are simply fine-tuned (but see~\cite{Senatore:2010wk,Barnaby:2011pe}).
In the case of supersymmetry, on the other hand, the observational prospects are considerably more optimistic.
Supersymmetry requires that the (real) inflaton field has a scalar partner in order to match the two degrees of freedom of the fermionic superpartner.
When SUSY is spontaneously broken by the spacetime curvature of the de Sitter background, this second scalar receives a mass of order $H$ (while the inflaton stays light either due to an additional global symmetry or due to some moderate fine-tuning).  This feature is special to SUSY, as internal symmetries cannot be spontaneously broken by curvature without violating the Coleman-Mandula theorem \cite{Coleman:1967ad}.  
We have proposed to use this massive scalar as a window into the supersymmetric origin of inflation.
Fortunately, the presence of a Hubble-mass scalar during inflation has a clean observational signature in the squeezed limit of the primordial bispectrum~\cite{Chen:2009zp}.
Notice that the path from naturalness via SUSY to Hubble-mass fields and the squeezed bispectrum requires little more than simply following your nose.

\vskip 4pt
An important by-product of this paper was the development of a general framework for systematically incorporating supersymmetry into the effective theory of inflationary fluctuations.
This supersymmetric effective theory of inflation has a wide range of applicability.
We presented two examples in detail: models with small sound speed and slow-roll quasi-single-field inflation. We showed how supersymmetry helps to address naturalness issues in a faithful way. 
It would be interesting to study the broader phenomenology of the SUSY EFT of inflation: models with higher derivatives, quasi-single-field inflation with small sound speed, etc.
Moreover, our treatment made some simplifying assumptions concerning the effects of SUSY breaking on the inflationary dynamics (see \S\ref{sec:break}; in particular, we chose separate multiplets for the SUSY breaking spurion $X$ and the inflaton $\Phi$.). We also restricted the size of superpotential terms and hence the couplings to the auxiliary supergravity fields. Finally, we didn't study scenarios in which the fluctuations in additional isocurvature fields are converted to curvature perturbations after inflation (see e.g.~\cite{Senatore:2010wk}).
Relaxing those assumptions will allow a broader class of models.
This may lead to additional observational signatures that haven't been anticipated yet.
In the meantime, we should also determine how present data constrains the models presented in this paper.
Efforts are under way to see what CMB and LSS data has to say about SUSY in the sky~\cite{Paper3}.

\acknowledgments
We are grateful to Xingang Chen, Thomas~Dumitrescu, Anatoly Dymarsky,  Guido~Festuccia, Eugene Lim, Marilena LoVerde, Liam McAllister, Marcel Schmittfull,
Nathan Seiberg, Leonardo Senatore, Kendrick Smith, and Matias Zaldarriaga for discussions.
D.B.~thanks the Institute for Advanced Study for hospitality and Dawn Dunbar for organizing the visit.
D.B.~gratefully acknowledges support from a Starting Grant of the European Research Council (ERC STG grant 279617) and partial support from STFC under grant ST/FOO2998/1.
The research of D.G.~is supported by the DOE under grant number DE-FG02-90ER40542 and the Martin A.~and Helen Chooljian Membership at the Institute for Advanced Study.  

\newpage
\appendix
\section{Supergravity for Effective Theories}
\label{sec:SUGRA}

In a companion paper \cite{Paper1}, we study how general effective theories with global supersymmetry are coupled to minimal supergravity.  In this appendix, we give a brief, and mostly qualitative, overview of that work.  

\vskip 4pt
The basic challenge in constructing a theory of supergravity arises from the fact that rigid SUSY is a spacetime symmetry.  Specifically, the SUSY algebra $\{ Q_{\alpha}, \bar Q_{\dalpha} \} =2 \sigma^{\mu}_{\alpha \dalpha} P_{\mu}$ relates SUSY transformations to spacetime translations.  In order to describe a supersymmetric theory on a curved background, one needs to promote both vector and spinor derivatives ($\partial_{\mu}$, $D_{\alpha}$, and $\Df_{\dalpha}$,) to covariant derivatives ($\nabla_{\mu}$, ${\cal D}_{\alpha}$ and $\Db_{\dalpha}$).  To do so, one promotes the metric and the spin connection to superfields, and imposes constraints and gauge-fixing conditions that reduce the number of new, off-shell, degrees of freedom to a manageable number.
Minimal supergravity represents one set of self-consistent constraints that allow us to define a covariant theory~\cite{WB}.  The off-shell description contains two dynamical fields, the vierbein $e_\mu^a$ and the gravitino $\psi_\mu^\alpha$, and two auxiliary fields, a complex scalar $M$ and a real vector $b_{\mu}$.  All geometric objects may be defined in terms of these fields.  
Superspace is then defined with new fermionic coordinates~$\Theta$.
With some work, the action in superspace may be written as
\beq\label{eqn:sugra}
S = \frac{1}{\kappa^2} \int \d^4 x \int d^2 \Theta \, E \left[ \tfrac{3}{8} (\Db^2 - 8 R) \, e^{-\tfrac{\kappa^2}{3}K(X, X^{\dagger},\Phi, \Phid, \partial_\mu \Phi, \partial_\mu \Phid,  \cdots )} + \kappa^2 W(X,\Phi)\right] + {\rm h.c.}\ ,
\eeq
where $\kappa \equiv 1/\Mp$.
There are two crucial elements that make this action manifestly supersymmetric: First, the operator $(\Db^2 - 8 R)$, with curvature superfield $R \supset -\tfrac{1}{6} M  +\tfrac{1}{12}  {\cal R} \Theta^2$, projects any scalar operator~$\mathcal{O}$ onto a chiral operator ${\cal C} = (\Db^2 - 8 R) \mathcal{O}$ (i.e~an operator satisfying $\Db_{\dalpha} {\cal C} = 0$).  Second, the chiral density, $E \supset {\rm det}\, e \,(1 -\Theta^2 \bar M)$, is defined such that $\int d^2 \Theta\, E \, {\cal C} $ is supersymmetric for any chiral operator ${\cal C}$.  

\vskip 4pt
In inflation, the most important supergravity contributions are those that arise from curvature couplings.  In minimal supergravity, the coefficients of the curvature couplings cannot be changed arbitrarily and they will induce masses and couplings that are controlled by the Hubble scale $H$.  For example, for slow-roll inflation, with $K_{\rm s.r.} = \tfrac{1}{2}(\Phi+\Phid)^2 + X X^\dagger$, eq.~(\ref{eqn:sugra}) leads to curvature couplings of the form
\beq\label{eqn:srcurvature}
\mathcal{L}_{\rm sugra}  \supset  - \tfrac{1}{2} \mathcal{R} \, \exp\left[-\tfrac{\kappa^2}{3} (\sigma^2 + |x|^2) \right] \ ,
\eeq
where ${\cal R} = -12H^2$ is the curvature of the de Sitter background.
When $K(\Phi, \Phid, X, X^{\dagger}, \ldots ) = K(\Phi, \Phid, X, X^{\dagger} ) \equiv K(\Phi_i, \Phi_i^{\dagger})$ (i.e.~when there are no higher-derivative terms or additional curvature couplings), the curvature coupling arises from a universal coupling to the chiral curvature superfield $R$.  In this case, one normally Weyl rescales the metric to go to Einstein frame and these curvature terms then appear as part of the supergravity scalar potential
\beq\label{eqn:scalar}
V(\phi_i)_{\rm sugra} = e^{\kappa^2 K(\phi_i, \phid_i)} \left(g^{i \bar{\jmath}} D_i W \overline{D_j W} - 3 \kappa^2 |W|^2 \right) \ ,
\eeq
where $D_i W \equiv \partial_{\phi_i} W + \kappa^2  (\partial_{\phi_i} K) W$ and $g^{i \bar{\jmath}}$ is the inverse of the K\"ahler metric.  This scalar potential holds for any two-derivative action.  
When $\langle \phi_i \rangle = 0$ for all $i$, the Weyl rescaling is unnecessary and we can determine the contributions to the scalar masses directly from the curvature coupling.  
Let us illustrate this for the case of slow-roll inflation: we may compute the mass of $\sigma$ directly from the F-term potential (\ref{eqn:scalar}) to find $m_{\sigma}^2 = 2\kappa^2 |{\cal F}_X|^2= 6 H^2$.  Alternatively, we can expand (\ref{eqn:srcurvature}) using ${\cal R} = -12 H^2$ to find one contribution to the mass of size $m^{2}_{\sigma (1)} = 4 H^2$.  By expanding the exponential (\ref{eqn:sugra}) to order $(\Phi + \Phi^{\dagger})^2 X^{\dagger} X$, we find an additional contribution, $m^{2}_{\sigma (2)} = 2 H^2$. Combining both terms, we find $m_{\sigma}^2 = m^{2}_{\sigma (1)} +m^{2}_{\sigma (2)} = 6 H^2$, as before.  

\vskip 4pt
Let us make a few comments on the role of the auxiliary fields $M$ and $b_\mu$:  in general, they can be integrated out to produce Planck-suppressed couplings between $X$ and $\Phi$.  For two-derivative actions, these effects have been included in writing the potential in eq.~(\ref{eqn:scalar}).  The contributions of these fields are only relevant when $M$ or $b_\mu$ acquire vevs.  However, as we explain in \S\ref{sec:sugra}, these vevs are small (or even strictly zero) during inflation and give at most subleading corrections to the masses of fields. 

\vskip 4pt
For two-derivative actions, the combined effects of curvature couplings, the auxiliary fields, and Planck-suppressed interactions, are all encoded in (\ref{eqn:scalar}).  However, for higher-derivative theories, there is no compact formula that includes all these terms~\cite{Paper1}. 
 In the context of inflation, determining the scalar potential remains relatively simple, as the auxiliary fields are negligible.  
 As a simple example, we studied theories with small sound speed, such as 
  \begin{align}
\label{equ:exampleX}
K_{c_s} 
\ =\ \tfrac{1}{2} (\G + {\G}^\dag)^2 \Big[ c_1+  \frac{c_2}{\Mp^2 |\dot H|}\, \partial_\mu  \G \partial^\mu {\G}^\dag \Big]  \ , 
\end{align}
where $c_1 \equiv   1 + \tfrac{1}{2} \big( 1- \tfrac{1}{c_s^2}\big)$ and $c_2 \equiv  \tfrac{1}{4} \big( 1 - \tfrac{1}{c_s^2}\big)$.
 In \cite{Paper1} we derive the component Lagrangian for this theory
 \beq
{\cal L}_{c_s} \ =\ c_1 {\cal L}_1 + \frac{c_2}{ \Mp^2 |\dot H|}\, {\cal L}_2 + \cdots\ ,
\eeq
where 
\begin{align}
{\cal L}_1 &= -  |\partial_\mu \phi|^2 + \left(\underline{\underline{\tfrac{1}{6} {\cal R} - \tfrac{\kappa^2}{3} |{\cal F}_X|^2}} \right) \sigma^2\ , \label{equ:L1xx} \\
{\cal L}_2 &= -  \left( |\partial_\mu \phi|^2 \right)^2 - 2 \partial_\mu \sigma \partial_\nu \sigma   \partial^\mu \phi \partial^\nu \phid \nonumber  \\
&\hspace{0.4cm}+ \sigma^2 \Big\{  \Big( \, \underline{\underline{\tfrac{1}{6} {\cal R} -  \tfrac{\kappa^2}{3} |{\cal F}_X|^2}} \, \Big) \big[ (\partial_\mu \varphi_c)^2 + (\partial_\mu \sigma)^2\big]  - \tfrac{1}{2} \underline{\underline{{\cal R}_{\mu \nu}}}  \big[ \partial^\mu \varphi_c \partial^\nu \varphi_c + \partial^\mu \sigma \partial^\nu \sigma\big] \nonumber \\
&\hspace{1.7cm}  + \tfrac{1}{2} \underline{\underline{\nabla_\mu \nabla_\nu}} \big[ \partial^\mu \varphi_c \partial^\nu \varphi_c + \partial^\mu \sigma \partial^\nu \sigma\big] +  \partial^\mu \phi \partial_\mu \Box \phid + \tfrac{1}{2}\Box|\partial_\mu \phi|^2 \Big\} \ . \label{equ:L2xx}
\end{align}
Using $\bar \varphi_c = (2\Mp^2 |\dot H|)^{1/2} t$ and ${\cal R}_{\mu \nu} = -3H^2 g_{\mu \nu}$ (de Sitter), we determine the mass of the partner of the inflaton, 
\beq
{\cal L}_{c_s} = \cdots +\left[ \, c_1\left(\tfrac{1}{6} {\cal R} - H^2 \right) + c_2 \left( \tfrac{1}{12}  {\cal R} - 2 H^2 \right) (\partial_\mu t)^2 + c_2 \nabla_\mu \nabla_\nu (\partial^\mu t \partial^\nu t) \right] \sigma^2  + \cdots \ .
\eeq
With $\nabla_\mu \nabla_\nu (\partial^\mu t \partial^\nu t) = 9H^2$, we find
\beq
m_\sigma^2 = 6H^2 c_1 -12 H^2 c_2 \simeq 0 \times \frac{H^2}{c_s^2} + 6 H^2\ . \label{equ:MSX}
\eeq
The cancellation of the leading term in (\ref{equ:MSX}) is completely accidental and not protected by any symmetry.  The mass received contributions both from curvature and from the exponentiation of the rigid action.  Neither the curvature couplings nor the form of the K\"ahler potential are fixed in minimal supergravity.  We could change either coefficient to eliminate the cancelation without breaking any symmetry of the action.  For example, there are contributions to the mass from modifications to the K\"ahler potential via Planck-suppressed couplings of the form
\beq
 \int d^2 \Theta\, E (\Db^2 -8 R )\frac{\beta}{\Mp^2} X^{\dag} X\, K_{c_s} \ \supset \ 3 \beta\, H^2 \sigma^2  \left[ c_1 +  \frac{c_2}{\Mp^2 |\dot H|}\, |\partial_\mu  \phi|^2 \right]\ .
\eeq
These contributions are model-dependent, but generically present.  
This contributes terms at the same order and change the numerical coefficient in eq.~(\ref{equ:MSX})
\beq
m_\sigma^2 \sim \frac{H^2}{c_s^2} \gg H^2\ .
\eeq
Hence, generically small sound speed will lead to a parametrically enhanced mass for $\sigma$.

\vskip 4pt
We refer the reader to
our companion paper~\cite{Paper1} for a more complete treatment of supergravity of more general effective theories.

\newpage
\section{Estimates for Quasi-Single-Field Inflation}
\label{sec:QSFI2}

In this appendix, we will explain how to estimate both the amplitude of the bispectrum and the scaling of the squeezed limit for models of quasi-single-field inflation.  In both cases, we will provide two different methods for estimating the results: 
\begin{enumerate}
\item[{\it i})] We will use dimensional analysis to estimate the answer from the full bispectrum calculation.  The advantage of this approach is that it essentially sets up the complete calculation and therefore there is little room for error in the estimate.  
\item[{\it ii})] We will estimate the answer directly based on physical reasoning.  The advantage of this approach is that it dramatically simplifies the process of estimating the results and provides a physical understanding of the origin of the effect.  
\end{enumerate}
Of course, both approaches will yield the same results.

\subsection{Amplitude of the Bispectrum}

We use the ``in-in formalism" (see e.g.~\cite{Weinberg:2005vy,Chen:2010xka}) to compute the bispectrum
\beq\label{eqn:inin}
\langle \zeta^3 (0) \rangle = \langle \Omega | \left[ \bar T {\rm exp}\Big( i \int_{-\infty}^{0} H_I (\tau) \d \tau \Big) \right] \zeta_{I}^3(0) \left[ T {\rm exp}\Big( - i \int_{-\infty}^{0} H_I (\tau) \d \tau \Big) \right]  | \Omega \rangle \ ,
\eeq
where the subscript $_I$ indicates interaction picture fields and $T$ denotes time ordering.  We can calculate the bispectrum of any operator in perturbation theory by expanding the time-ordered exponentials to the appropriate order in the interaction Hamiltonian $H_I$.

\vskip 4pt
{\it Direct estimation.}\hskip 6pt  As a concrete example, let us consider the original model of QSFI~\cite{Chen:2009zp}, with interaction $\int \d^4 x \sqrt{-g} \, \mu \sigma^3$ and mixing term $\int \d^4 x \sqrt{-g} \, m^3 \dot \pi \sigma$.
The bispectrum generated by these interactions is schematically of the form
\beq
\includegraphicsbox[scale=.7]{QSFI_feynman3}  \qquad \Leftrightarrow \qquad  \langle \zeta_k^3 \rangle = \zeta_k(0)^3\, \prod_{i=1}^3 \int \d \tau_i\, a^3 m^3  \pi' \sigma(\tau_i) \cdot \int \d\tau \, a^4 \mu \sigma^3 (\tau) \ ,
\eeq
where $'$ denotes a derivative with respect to conformal time $\tau$.  Here, we have ignored the details of the time ordering and the commutator structure that follows from eq.~(\ref{eqn:inin}).  We will be estimating the results by dimensional analysis of the integrals which will be insensitive to these subtleties---although these subtleties can be important to prove convergence of the integrals~\cite{Chen:2009zp}.
To estimate the integrals, we will need the mode functions for $\pi$ and $\sigma$.  Since the Goldstone boson is massless in the decoupling limit, its mode function in de Sitter space is 
\beq
\label{equ:piM}
\pi_k (\tau) = \frac{1}{\sqrt{2 k^2}} \frac{H}{(\Mp^2 |\dot H|)^{1/2}} (1+ i k \tau) e^{-i k \tau} \ .
\eeq
The mode functions for the massive field $\sigma$ are Hankel functions.  When $|k\tau| \ll 1$, these simplify to
\beq
\label{equ:sigmaM}
\lim_{k\tau \to 0}\sigma_k = H \frac{(-\tau)^{3/2 - \nu}}{k^{\nu}} , \qquad {\rm where} \qquad \nu \equiv \sqrt{\frac{9}{4} - \frac{m^2_\sigma}{H^2} } \ .
\eeq
This long-wavelength behavior will be sufficient for estimating the integrals, since the modes oscillate rapidly and suppress the integrals when $|k \tau| \gg 1$.

The first integral that we wish to estimate is the mixing term $\int \d \tau\, a^3 m^3  \pi' \sigma$.  Using eqs.~(\ref{equ:piM}) and (\ref{equ:sigmaM}), we can write this as
\beq
\int^{\tau_\star} \d \tau\, a^3 m^3  \pi' \sigma \sim \rho \int^{\tau_\star} \d \tau\, \frac{H^2}{(-H \tau)^3}  \frac{k^2 \tau}{\sqrt{2 k^3}} e^{- i k \tau} \frac{(-\tau)^{3/2 - \nu}}{k^{\nu}} \sim \frac{\rho}{H} \int^{\tau_\star} \d \tau\, k^{1/2 - \nu} \tau^{-\nu-1/2} e^{- i k \tau} \ . \label{equ:INT}
\eeq
As we stated above, when $|k \tau| \gg 1$, the integral is exponentially suppressed by the rapid oscillations of the mode functions.  Since the only scale in the problem is $k$, the integral will be dominated by $\tau \sim k^{-1}$.  For large values of $\nu$ one might worry that the integral is actually divergent as the upper limit of integration $\tau_\star \to 0$, but a more careful analysis shows that all power law divergences cancel~\cite{Chen:2009zp}.  In the following, we will take it as a given that there aren't any power law divergences. This ensures that the integral receives its dominant contribution from $\tau \sim k^{-1}$.  Moreover, this observation also shows that these integrals are only weakly dependent on the upper limit of integration.  The integral in (\ref{equ:INT}) may then be estimated as
\beq
\int \d \tau\, a^3 m^3  \pi' \sigma \sim\frac{\rho}{H} k^0 \ .
\eeq
This leads to the ``Feynman rule" that each insertion of the mixing contributes a momentum-independent factor of $\rho/H$ to the bispectrum amplitude, $f_{\rm NL}$.

The interaction term, $\int \d\tau\, a^4 \mu \sigma^3$, can be estimated in a similar way,
\beq
\int \d \tau\, a^4 \mu \sigma^3 \sim \mu \int \d \tau\, \frac{H^3}{(-H \tau)^4} \frac{(-\tau)^{9/2 - 3\nu}}{k^{3\nu}}  \sim \frac{\mu}{H} \int \d\tau\, \frac{\tau^{1/2 - 3 \nu}}{k^{3 \nu}} \sim \frac{\mu}{H}\, k^{-3/2} \ .
\eeq
We have again assumed that any power law divergences cancel and that the integral is dominated by $\tau \sim k^{-1}$.  
We see that each interaction vertex leads to a factor of $\mu/H$.

Putting these estimates together, we find
\beq
\langle \zeta_k^3 \rangle \sim  \frac{\Delta_\zeta^4}{k^6} \left(\frac{\rho^3}{H^3} \frac{\mu}{H} \Delta_\zeta^{-1} \right)  \qquad \Rightarrow \qquad f_{\rm NL} \sim \frac{\rho^3}{H^3} \frac{\mu}{H} \Delta_\zeta^{-1}  \ , \label{equ:careful}
\eeq
where $\Delta_\zeta^2 \equiv H^4/(4\Mp^2 |\dot H|)$.
This result is consistent with the complete calculation in \cite{Chen:2009zp}.

\vskip 4pt
{\it Quick estimate.} \hskip 6pt 
While the estimate that we just performed is robust, it relies on actually setting up the full calculation.  
We would like to have a faster and possibly more intuitive explanation for the result.  In single-field models, we can often estimate the size of non-Gaussianity as
\beq
f_{\rm NL}  \sim \frac{1}{\zeta} \frac{\mathcal{L}_3}{\mathcal{L}_2}\ , \label{equ:FNL}
\eeq
 where the r.h.s is evaluated at horizon crossing.
At first glance, it may not be clear how such an estimate would work for quasi-single-field inflation.  First of all, the interaction $\mathcal{L}_3$ involves only $\sigma$, while the power spectrum involves just $\pi$.  This problem is easily resolved by using the fact that the mixing term is independent of $k$.  Specifically, the mixing term allows us to contract $\pi$ with $\sigma$ at the cost of a numerical suppression, $\rho / H$.  
In principle, the mixing term therefore allows us to compute the ratio in (\ref{equ:FNL}).  The second issue we face is that the $\sigma$ modes evolve outside the horizon.  To make an estimate, we therefore need to specify the time at which the amplitude of $\sigma$ is measured.  However, our previous estimates have shown that the dominant contributions to the integrals comes from the time of horizon crossing $\tau \sim k^{-1}$.  Since $f_{\rm NL}$ is defined as the amplitude of the bispectrum in the equilateral configuration ($k_1 =k_2 =k_3$), we may evaluated each $\sigma$ at horizon crossing.  This allow us to rewrite $\sigma$ in term of $\pi$,
\beq
\sigma_{k \sim 1/\tau} \ \to\ \frac{\rho}{H} (\Mp^2 |\dot H|)^{1/2} \, \pi_{k \sim 1/\tau} = \frac{\rho}{H} \frac{H^2}{\Delta_\zeta} \, \pi_{k \sim 1/\tau}  \ . \label{equ:enh}
\eeq
We can then estimate $f_{\rm NL}$ from eq.~(\ref{equ:FNL}), 
\beq\label{equ:estimatea}
f_{\rm NL}  \sim  \frac{1}{\zeta} \frac{\mu \sigma^3}{\Mp^2 \dot H (\partial_\mu \pi)^2} \sim
 \frac{\rho^3}{H^3} \frac{\mu}{H} \Delta_\zeta^{-1} \ .
\eeq
We see that this estimate matches the more careful estimate in eq.~(\ref{equ:careful}).

\vskip 4pt
{\it Generalizations.} \hskip 6pt  Above we reproduced the result for the specific QSFI example of \cite{Chen:2009zp}. 
In Section~\ref{sec:QSFI}, we explored variations of QSFI, allowing for all possible interactions between $\pi$ and $\sigma$ consistent with the symmetries of the effective theory of inflation. 
We now have the tools to understand the non-Gaussianity in all these examples and to recover the results of Table~\ref{tab:NG}:

First, we consider the alternative mixing term of the form 
\beq 
\alpha (\Mp^2 |\dot H|)^{1/2} \dot \pi \dot \sigma\ .
\eeq  
By repeating the above analysis, we find that the integral associated with these mixing terms scales as $\alpha k^0$.  
The coupling $\alpha$ therefore now plays the role of $\rho/H$. The rest is unchanged.
Hence, replacing $\rho/H$ by $\alpha$ gives the result for the new mixing terms.

Next, we consider models with different interactions, i.e.~different couplings between $\pi$ and $\sigma$ and different derivative structures. Applying the methods of this appendix, we can determine Table~\ref{tab:NG}. We find an enhancement in $f_{\rm NL}$ by factors of $\Delta_{\zeta}^{-1}$ for certain cubic interactions. This can be explained by the larger amplitude of $\sigma$ fluctuations at horizon crossing relative to the fluctuations of $\pi$. From eq.~(\ref{equ:enh}) we see that we gain by powers of $\Delta_{\zeta}^{-1}$ for every factor of $\sigma$ that appears in $\mathcal{L}_3$.  For example, estimating $f_{\rm NL}$ for the interaction $\bar m_a \partial_\mu \pi \partial^\mu \sigma \sigma$, we find
\beq
f_{\rm NL} \sim \frac{1}{\zeta} \frac{\mathcal{L}_3}{\mathcal{L}_2} = \frac{1}{\zeta}\frac{\bar m_a \partial_\mu \pi \partial^\mu \sigma  \sigma}{\Mp^2 \dot H (\partial_\mu \pi)^2} \sim \frac{\bar m_a \pi}{\zeta} \frac{\rho^2}{H^2} = \frac{\bar m_a}{H}\frac{\rho^2}{H^2} \ .
\eeq
As expected, $f_{\rm NL}$ for this interaction is suppressed relative to (\ref{equ:estimatea}) by one power of $\Delta_\zeta$.

\subsection{Squeezed Limit of the Bispectrum}

In the previous section, we showed how to estimate the amplitude of the bispectrum.  We will now use a similar analysis to determine its scaling behavior in the squeezed limit.

\vskip 4pt
{\it Direct estimation.} \hskip 6pt  The most reliable way to estimate the squeezed limit is by dimensional analysis.  We again start with the schematic form of the bispectrum
\beq
\langle \zeta_{{\bf k}_1}\zeta_{{\bf k}_2} \zeta_{{\bf k}_3} \rangle (0) =\delta({\bf k}_1+ {\bf k}_2 + {\bf k}_3) \left[ \zeta_{k_1}\zeta_{k_2} \zeta_{k_3}(0) \cdot \prod_{i=1}^3 \int \d \tau_i\, a^3 m^3  \pi_{k_i}' \sigma_{k_i} \cdot \int \d\tau \, a^4 \mu\, \sigma_{k_1} \sigma_{k_2}\sigma_{k_3} \right] \ , \label{equ:EST}
\eeq
where the delta function enforces $k_2 = k_3$ in the squeezed limit, $k_1 \to 0$.
As before, we will estimate the behavior of each piece separately:

The easiest piece to understand is the contribution from the overall factor $\zeta_{k_1}\zeta_{k_2} \zeta_{k_3}(0)$, which scales as
\beq
 \zeta_{k_1}\zeta_{k_2} \zeta_{k_3}(0) =  \frac{\Delta_\zeta^{3}}{k_1^{3/2} k_2^3} \ .
\eeq

Next, we consider the mixing terms. As we saw in the previous section, they scale as $k^0$.  In principle, they could therefore contribute to the squeezed limit through ratios like $(k_1 / k_2)^n$.  However, this does not occur, since each mixing term is only a function of a single momentum.  Specifically, the integral over the long-wavelength modes with momentum $k_1$ takes the form
\beq
\int^{\tau_\star} \d \tau\, a^3 m^3  \pi_{k_1}' \sigma_{k_1}  \sim \frac{\rho}{H} \int^{\tau_\star} \d \tau\, k_1^{1/2 - \nu} \tau^{-\nu-1/2} e^{- i k_1 \tau} \sim \frac{\rho}{H} \ .
\eeq
As before, we estimated the integral by taking the dominant contribution to the integral to come from $\tau \sim k_1^{-1}$.  The only way the scale $k_2$ could have entered the estimate, is through an implicit $k_2$-dependence in the integration limit $\tau_\star$.  However, as we explained above, the cancelation of divergences as $\tau_\star \to 0$, also implies that the integral is only weakly dependent on the limits of integration.  As a result, any $k_2$-dependence from the limit of integration is a subleading effect. The skeptical reader is encouraged to consult the full in-in calculation~\cite{Chen:2009zp} to confirm this.

The final contribution is from the interaction term.  In the previous section, we found that this integral scales as $k^{-3/2}$, as required for scale invariance of the bispectrum.  Repeating this estimate, but now being careful to distinguish the long mode, $k_1$, from the short modes, $k_2$, we find
\beq\label{eqn:squeezedint}
\lim_{k_1 \to 0} \int \d \tau\, \mu \sigma_{k_1} \sigma_{k_2} \sigma_{k_3}  \sim \frac{\mu}{H} \int \d\tau\, \frac{\tau^{1/2 - 3 \nu}}{k_1^{\nu}k_2^{ 2\nu}} \ .
\eeq
The integral contains two scales, $k_1$ and $k_2$, so it may not be obvious which scale dominates.  When $|\tau| \sim k_1^{-1} \gg k_2^{-1}$, the short modes are still well inside the horizon and oscillate rapidly.  Therefore, the integral is exponentially suppressed (this isn't obvious in eq.~(\ref{eqn:squeezedint}) because we have taken the mode functions in the limit where the modes are outside the horizon).  It isn't until the short modes freeze out, $|\tau| \sim k_2^{-1}$, that we get a significant contribution to the integral.  Therefore, we can estimate the integral as if there were only one scale, $k_2$, to get
\beq
\lim_{k_1 \to 0} \int \d \tau\, \mu \sigma_{k_1} \sigma_{k_2} \sigma_{k_3}  \sim \frac{\mu}{H} \frac{1}{k_1^{\nu} k_2^{3/2 - \nu}} \ .
\eeq

Putting all these estimates together, we find
\beq
\lim_{k_1 \to 0} \langle \zeta_{k_1}\zeta_{k_2} \zeta_{k_2} \rangle \, \propto\, \frac{1}{k_1^{3/2 + \nu} k_2^{9/2 - \nu}} \ .
\eeq
This agrees with the result of \cite{Chen:2009zp}.

\vskip 4pt
{\it Intuitive explanation.} \hskip 6pt  
To gain some intuitive understanding for the scaling of the bispectrum in the squeezed limit, it is useful to rewrite the result as
\beq
\lim_{k_1 \to 0} \langle \zeta_{{\bf k}_1} \zeta_{{\bf k}_2} \zeta_{{\bf k}_3}\rangle \, \propto\, \underbrace{\frac{1}{k_1^3 k_2^3}}_{\rm local} \left( \frac{k_1}{k_2}\right)^{{3/2} -\nu} \ . \label{equ:SQUloc}
\eeq
The first term has the same squeezed limit as the {\it local shape}, $k_1^{-3}$.  
The second term modifies this scaling by the ratio $(k_1 / k_2)^{3/2 - \nu}$. For $\nu < 3/2$ (or $m_\sigma >0$), this leads to a suppression in the squeezed limit, while for $\nu =3/2$ (or $m_\sigma =0$) we recover the local shape.
This suggests that the physics of the squeezed limit of QSFI is the same as the local shape, with a modification induced by the superhorizon behavior of the mode function for the massive field~$\sigma$.

Let us try to understand this in a bit more detail.
The squeezed limit of the local shape can be understood as a local modulation of the short-scale power due to the presence of the long-wavelength mode.  For $m_\sigma^2 = 0$ in QSFI, we recover the same result.  On the other hand, when $m_\sigma^2 > 0$, the isocurvaton decays outside the horizon.  Although the modulation of the power spectrum is still local in space, the amplitude of the long mode now depends on the time of horizon crossing.  Specifically, the interaction plays an important role at horizon crossing of the short modes, $k_2$.  When the short modes cross the horizon, $|k_2 \tau_2| \sim 1$, the amplitude of the long mode, $k_1$, is suppressed relative to its amplitude at its own horizon crossing, $|k_1 \tau_1| \sim 1$,
\beq
\sigma_{k_1}(\tau_2) = \sigma_{k_1}(\tau_1 )\left( \frac{k_1}{k_2}\right)^{{3/2} -\nu} \ .
\eeq
We hence understand that the deviation from the local shape in (\ref{equ:SQUloc}) is the result of the decay of the long-wavelength mode between the time it crosses the horizon and the time the short modes cross the horizon.
The mixing term also plays an important role in this interpretation.
It allows the conversion of the massive mode~$\sigma$ into the massless mode $\zeta = - H \pi$ shortly after horizon crossing.  If one simply computed the bispectrum for $\sigma$ using the cubic interaction, one wouldn't find the same squeezed limit.  Unless the $\sigma$ fluctuations are converted into a massless mode at horizon crossing, they will continue to decay outside the horizon.  This would lead to additional factors of $k_1$ and $k_2$ related to the decay of the amplitude from horizon crossing until the end of inflation.  

\vskip 4pt
{\it Generalizations.} \hskip 6pt The behavior in the squeezed limit is robust to changes of the model: 

 The specific form of the mixing term is unimportant, as its only role in life is to convert a massive mode into a massless mode.  As we discussed in the previous section, even when we include more derivatives in the mixing term (e.g.~$\dot \pi \dot \sigma$), the integral still scales like $k^0$.  As a result, the estimate of the mixing integral is the same in all cases and does affect the squeezed limit.
Changes to the interaction also do not alter the squeezed limit, as long as there is at least one field with no derivatives acting on it.  The behavior in the squeezed limit, is the result of a ``local" correlation of the long and short modes.  When there are no derivatives acting on the field, the short modes are sensitive only to the amplitude of the long mode, at the time of horizon crossing.  In the case, where all fields are acted on by derivatives, the short modes are only sensitive to gradients of the long mode, and therefore the squeezed limit is further suppressed by a factor of~$k_1^2$.

\newpage
 \begingroup\raggedright\endgroup

\end{document}